\documentstyle[a4wide,12pt]{article}

\begin{document}

\vspace{1 cm}

\begin{center}

{\huge {\bf Quark-Antiquark Bound States and the Breit Equation}}\\

\vspace{1 cm}  

George D. Tsibidis  \\
{\it School of Electrical and Electronic Engineering, \\
University of Birmingham, \\Edgbaston, Birmingham, B15 2TT, U.K}\\
E-Mail address: georgiot@eee-fs7.bham.ac.uk\\

\end{center}

\vspace{1cm}

\begin{abstract}

A non-covariant but approximately relativistic two-body wave equation 
(Breit equation) describing the quantum mechanics 
of two fermions interacting with 
one another through a potential containing scalar, pseudoscalar and
vector parts is presented. 
After expressing the sixteen component two-body
wavefunction in terms of a radial and an angular function by means of the
multipole expansion, the initial
equation can be reduced into a set of sixteen radial equations which, in
turn, can be classified in accordance to the parity and the state 
of the wavefunctions involved. The adequacy of the reduced 
equations in describing real problems is discussed by 
applying the theory to QCD problems and the calculation of the energies 
of bound states of quark-antiquark systems is performed 
to order $\alpha^4$.    
We show that bound states of heavy quarks can be described adequately 
by the Breit equation for a funnel interaction between the particles.\\ 
{\it PACS numbers}: 03.65.Ge; 03.65.Pm; 12.39.Jh; 12.39.Ki. \\

\end{abstract}

\baselineskip=20pt 

\newpage

\setcounter{page}{1} 

\pagestyle{plain}

\section{Introduction}

Over the years a number of equations has been introduced in an effort to
describe adequately the relativistic dynamics of a system of two
interacting spin-$\frac 12$ particles (\cite{a1}-\cite{a6}). The problem 
is to determine how the
two fermions behave when they are influenced by their mutual interaction.
The simplest relativistic equation for fermions, the Dirac equation, which
can describe the quantum mechanics of a single fermion, is not useful for a
two-fermion system unless the mass of one of the particles is much larger
than that of the other particle. One equation that has been extensively used, 
in the past, is the equation Breit proposed in 1929, which 
describes the interaction between two electrons 
\cite{a2,a6,a7,a8}
\begin{equation}
[E-H_{(1)}-H_{(2)}-V_{int}(|\mbox{\boldmath $\vec{x}_1$}-%
\mbox{\boldmath $\vec{x}_2$}|)\psi (\mbox{\boldmath $\vec{x}_1$},%
\mbox{\boldmath $\vec{x}_2$})]=0 				
\label{eq:breit}
\end{equation}
where $H_{(i)}=\mbox{\boldmath $\vec{a}_i$}\cdot 
\mbox{\boldmath
$\vec{p}_i$}+\mbox{$\beta$}m$ ({\it i}=1,2) is identical with the Dirac
Hamiltonian of the {\it i} particle and {\it E} is the total energy of the
system. The interaction between the two particles equals 
$V_{int}=V_c+V_B$ and it 
is partly due 
to an {\it instantenous} (static) Coulomb interaction $V_c$ and partly 
due to effects prescribed by quantum electrodynamics described by 
{\it retardation } terms 
$$
V_B(\mbox{\boldmath $\vec{r}$})=-\frac 12V_c(\mbox{\boldmath $\vec{r}$})(%
\mbox{\boldmath
$\vec{a}$}_1\cdot \mbox{\boldmath
$\vec{a}$}_2+\mbox{\boldmath
$\vec{a}$}_1\cdot \mbox{\boldmath $\widehat{r}$}\mbox{\boldmath
$\vec{a}$}_2\cdot \mbox{\boldmath $\widehat{r}$}) 
$$
which is referred to, as ``Breit Interaction''. 
$\mbox{\boldmath $\vec{a}$}_i$ are the Dirac matrices, 
$\mbox{\boldmath $\vec{r}$}(\equiv \mbox{\boldmath
$\vec{x}$}_1-\mbox{\boldmath $\vec{x}$}_2)$ is the relative position 
vector and $\mbox{\boldmath $\widehat{r}$}$ is the unit vector.

Although the initial Breit equation involved two 
electrons \cite{a7}, we aim to 
generalise it for any two fermions. 
There are two requirements that the Eq.(\ref{eq:breit}) 
clearly satisfies: (i) In the
limit of negligent interaction between the particles, 
Eq.(\ref{eq:breit}) implies that
the total energy of the system equals the sum of the energies of the
particles and the stationary wavefunction $\psi (%
\mbox{\boldmath
$\vec{x}_1$},\mbox{\boldmath $\vec{x}_2$})$ is the product of the
wavefunctions $\psi (\mbox{\boldmath $\vec{x}_1$})$, $\psi (%
\mbox{\boldmath
$\vec{x}_2$})$ which are simply the solutions of the Lorentz-invariant Dirac
equations for each particle, separately, (ii) In the non-relativistic limit,
the Breit equation reduces to the Schr\"odinger equation for a two-particle
system.

If retardation effects are ignored, the Breit equation 
reduces to the Dirac equation in the infinite limit of the mass of one of the 
constituent particles and then, either equation can be
used. Certainly, the former 
would provide a more accurate description because corrections due to the 
motion of the heavier particle should be taken into account.

Unlike the Dirac equation, the Breit equation lacks a very important
ingredient characterising the relativistic equations, 
the covariance. Eq.(\ref{eq:breit}) is not written in 
a covariant notation since individual terms in the part 
representing the interaction are not Lorentz invariant (the potential is 
not a relativistically invariant quantity). Besides 
this, not only does the wavefunction in the Breit equation 
depend on the positions $\mbox{\boldmath $\vec{x}$}_1$ and 
$\mbox{\boldmath $\vec{x}$}_2$ but also depends on one time-variable 
(rather 
than an one
time-variable each for the two particles), a fact that does not allow a
Lorentz-invariant formulation. However, it can be considered covariant in
the centre-of-mass-frame because, in that case, the relative motion of the
particles is studied and there is only one position vector which comes into
the equation, namely that one which measures the distance between the two
electrons, $\mbox{\boldmath$\vec{r}$}$. 
Attempts have been made aiming at deriving a covariant Breit equation by
means of constraining the equations. It is too early to say whether this
work is successful \cite{a9}.

In spite of the fact that Breit equation manifestly lacks covariance, 
it appears that it can describe
efficiently the quantum mechanics of some two-body systems (at least
approximately to the required accuracy) for long-range interactions with
small coupling \cite{a7,a10}. When the equation was proposed by Breit 
in 1929 \cite{a2}, QED was known
and the long range Coulomb type potential $V_c(\mbox{\boldmath
$\vec{r}$})$ was the form of the instantaneous
interaction for hydrogen-like atoms. In contrast to the 
static $V_c(\mbox{\boldmath $\vec{r}$})$ which is the
zero-order term in the {\it (v/c) } expansion 
of the electromagnetic interaction, the
Breit interaction $V_B(\mbox{\boldmath $\vec{r}$})$ is the {\it (v/c)}$^2$
term (velocity-dependent) and it constitutes only an approximation to
the relativistic interaction between the two particles. The study of the fine and hyperfine
structure for hydrogen-like atoms indicates that the Breit interaction should
not be considered on the same footing as the Coulomb potential $V_c(%
\mbox{\boldmath $\vec{r}$})$. On the 
contrary, it should be treated as a
small perturbation, otherwise it does not lead to correct 
results (\cite{a10}-\cite{a15}).  
The fact that 
the Breit equation gives satisfactory results for 
the Coulomb potential implies that the equation, 
although not covariant, can provide a good description 
to a two-body system for long-range interactions.   

In this work, we aim to extend the application of the 
Breit equation to systems of fermions that interact 
with one another through a short-range strong potential. 
Due to the fact that the distance between the two particles is 
very small, one position vector suffices to 
describe the system which puts in the same footing time and position. 
Then, one might consider the Breit equation is 
approximately covariant and more compatible with the 
special theory of relativity, in the case of short-range 
strong interactions. 

There has been a belief that the internal 
dynamics of quark-antiquark systems can be described satisfactorily 
by two-body semi-relativistic equations and some attempts have been made 
towards this direction (\cite{a13},\cite{b1}-\cite{b35}). This is a very 
interesting possibility since it suggests a 
$q \overline{q}$ bound state could provide a sort of ``Hydrogen-atom'' 
for QCD. In this paper, we consider fermionia composed from heavy 
quarks (bottom, charm) and we assume that 
the particles interact with each other through a 
``funnel'' potential \cite{b1,b20}. Throughout this paper, 
we will handle the Breit equation by taking into account 
the instantaneous interaction between the particles and 
treating the non-static terms perturbatively. It turns out that the 
Breit equation offers a very good desciption of the systems and this 
is partly due to the fact that the consituents are heavy particles 
whose speed are very small so as to declare that 
a semi-relativistic treatment gives satisfactory results 
for certain bound states of the two fermions. 
Unfortunately, this does not seem to hold true in the event of 
higher bound states as well as for bound states of lighter particles
because a relativistic treatment is necessary.
       
One of the advantages of the Breit-equation is that it can be handled
relatively easily by means of some standard techniques (i.e. multipole
technique \cite{b10,a16,a17,a18,a20}). The wavefunction describing the system is a sixteen component
function and it can be used to obtain radial equations. 

This paper is organised as follows. We start in Section \ref{sec:two} 
with the introduction
of the Breit equation and we assume a general interaction between two
spin-1/2 particles including a vector, scalar and pseudoscalar 
particle exchange. Using the multipole technique, we separate the equation
into its radial and angular parts and derive sixteen radial equations
grouped according to the values of the spin of the system. Then, 
we present the non-relavistic type of the potential which describes 
the interaction between two quarks [Section \ref{sec:three}] and we derive 
the radial equations that apply in that case [Section \ref{sec:four}]. In 
Section \ref{sec:five}, we solve the equations for two types of 
quarkonia, bottomium and charmonium, and we calculate the energy levels of 
various bound states. 
This consideration allows us to check the validity of the Breit equation 
by comparing the results derived from the theory with well-established 
results for bottomium and charmonium. Section \ref{sec:ten} contains brief concluding remarks.


\section{The Breit equation and its reduction to radial equations}
\label{sec:two}

We consider a two-body Dirac equation (Breit equation) 
\cite{a2,a6,a7,a8} 
\begin{equation}
\left[ E-\gamma _0^{(1)}(\mbox{\boldmath $\vec{\gamma}$}^{(1)}\cdot 
\mbox{\boldmath
$\vec{p}$}+m_1)-\gamma _0^{(2)}(-\mbox{\boldmath $\vec{\gamma}$}^{(2)}\cdot 
\mbox{\boldmath $\vec{p}$}+m_2)-V_{int}(|\mbox{\boldmath
$\vec{r}$}|)\right] \psi (\mbox{\boldmath $\vec{r}$})=0 
\label{eq:bre2}
\end{equation}
describing a system of two spin-$1/2$ particles of mases $m_1$ and $m_2$%
, in the centre of mass frame, interacting with each other 
through a static central potential of the form \cite{a10,a20}
\begin{eqnarray}
V_{int}(r)&=&V_S(r)+V_P(r)+V_V(r)\\\nonumber
V_S(r)&=&-(\gamma^{(1)}_{0}\otimes\gamma^{(2)}_{0})S(r) \\\nonumber
V_P(r)&=&-(i\gamma^{(1)}_{0}\gamma^{(1)}_{5})\otimes(i\gamma^{(2)}_{0}
\gamma^{(2)}_{5})P(r) \\ \nonumber
V_V(r)&=&+{[(\gamma^{(1)}_{0}\gamma^{(1)}_{\mu})\otimes
(\gamma^{(2)}_{0}\gamma^{(2)}_{\mu})]}_{\mu=0}V(r)\nonumber 
\label{bre:int}
\end{eqnarray}
where $\mbox{\boldmath $\vec{r}$}\equiv \mbox{\boldmath
$\vec{x}$}_1-\mbox{\boldmath $\vec{x}$}_2$, $\mbox{\boldmath $\vec{p}$}%
\equiv \mbox{\boldmath
$\vec{p}$}_1=-\mbox{\boldmath $\vec{p}$}_2$, $r\equiv |%
\mbox{\boldmath
$\vec{x}$}_1-\mbox{\boldmath $\vec{x}$}_2|$, and $V_S(r)$, $V_P(r)$ and $%
V_V(r)$ are the parts of the interaction with scalar, pseudoscalar and
vector Lorentz structure, respectively. 
The choice of the combination of the $\gamma $
matrices which leads to the appropriate Lorentz structure is not unique,
however, the selection is based on some conditions which should be satisfied.
To be more specific, $\gamma _0^{(1)}+\gamma
_0^{(2)}$ has a scalar Lorentz structure, as well, and although it seems
reasonable because it couples the potential $S(r)$ directly to the mass of
each particle (a feature of the scalar potential in contrast to the vector
potential which couples to the charge of the particles) does not flip
the helicities of the fermions in the case of the chiral representation.\\
The signs in front of the various forms of the potentials 
can be justified as follows: in
QED, the vector exchange is actually the Coulomb interaction between
particles and if their charges have the same sign, then $V_V(r)$ is
positive. On the other hand, for scalar and pseudoscalar exchange the
sign of propagator is opposite to the Coulomb part of the photon propagator.\\ The
superscript $i\;(i=1,2)$ which appears in the $\gamma $ matrices refers to
the particle. The reason why in the case of potentials with vector
Lorentz structure only the $\gamma _0^{(1)}$, $\gamma _0^{(2)}$
contributions have been taken into account, is that the $%
\mbox{\boldmath $\vec{\gamma}$}^{(i)}$ matrices give a non-static character
to the potential since they introduce velocity terms \footnote{%
We recall the relation $\mbox{\boldmath
$\vec{\gamma}$}^{(i)}=\gamma^{(i)}_0\mbox{\boldmath
$\vec{\alpha}$}^{(i)}=\gamma^{(i)}_0 \frac{\mbox{\boldmath $\vec{v}$}^{(i)}}{%
c}$ , where $\mbox{\boldmath $\vec{v}$}^{(i)}$ is the velocity of the {\it i}
particle.}. $\gamma ^{(1)}(\gamma ^{(2)})$ is the Dirac matrix $\gamma $
acting in the subspace of the spinor of particle 1 (2) and it acts on $\psi $
from the left (right) 
\begin{eqnarray}
\begin{array}[]{ccl}
\gamma^{(1)}\psi &\equiv& \gamma^{(1)}\psi\\
\gamma^{(2)}\psi &\equiv& \psi(\gamma^{(2)})^T
\end{array}
\label{gam:ma}
\end{eqnarray}
The total spin of the system of the two particles is either $S=0$ or $S=1$, 
therefore the total angular momentum of the system is either $j=l$ (for $S=0$%
) or $j=l\pm 1,l$ (for $S=1$).

The form Eq.(\ref{eq:bre2}) 
acquires depends on the representation of the $\gamma $
matrices. In the next subsection, Eq.(\ref{eq:bre2}) is writen in 
the Dirac-Pauli representation, where $\gamma _0^{(i)}$ 
matrices are taken to be diagonal. This choice will allow us 
to examine the connection with the non-relativistic limit.

\subsection{Dirac representation}

In the Dirac representation, $\gamma _0^{(i)}$ matrix is diagonal 
\begin{equation}
\gamma _0=\beta =\left( 
\begin{array}{cc}
{\bf 1} & 0 \\ 
0 & {\bf -1}
\end{array}
\right) ,\gamma _5=\left( 
\begin{array}{cc}
0 & {\bf 1} \\ {\bf 1} & 0
\end{array}
\right) ,\mbox {\boldmath $\vec \gamma $} =\left( 
\begin{array}{cc}
0 & \mbox {\boldmath $\vec {{\ \sigma }}$} \\ -\mbox {\boldmath $\vec {{\ \sigma }}$} & 0
\end{array}
\right) 
\label{gam:ma2}
\end{equation}
The spinor $\psi (\mbox{\boldmath $\vec{r}$})$ is a sixteen-component wave
function and it can be represented as a $4\times 4$ matrix 
\begin{equation}
\psi (\mbox{\boldmath $\vec r$})=(\psi _{\gamma _0^{(1)}\gamma _0^{(2)}})=\left( 
\begin{array}{cc}
\psi _{++} & \psi _{+-} \\ 
\psi _{-+} & \psi _{--}
\end{array}
\right) 
\label{wav:fun}
\end{equation}
where the indices $+,-$ are the eigenvalues ($+$1, $-$1) of the Dirac
matrices $\gamma _0^{(1)},\gamma _0^{(2)}$ in the so-called double Dirac
representation \cite{a10,a17,a18}. The left index refers to the first particle and the right
one to the second particle. By inserting (\ref{bre:int}) and 
(\ref{gam:ma2}) and (\ref{wav:fun}) into Eq.(\ref{eq:bre2}), the Breit 
equation takes the form 
\begin{eqnarray}
&E&\left(
\begin{array}{cc}
\psi _{++} & \psi _{+-} \\
\psi _{-+} & \psi _{--}
\end{array}
\right)-
\left(
\begin{array}{cc}
\mbox{\boldmath $\vec{\sigma}$}^{(1)}\cdot
\mbox{\boldmath
$\vec{p}$}\psi _{-+} & \mbox{\boldmath $\vec{\sigma}$}^{(1)}\cdot  
\mbox{\boldmath
$\vec{p}$}\psi _{--} \\
\mbox{\boldmath $\vec{\sigma}$}^{(1)}\cdot  
\mbox{\boldmath
$\vec{p}$}\psi _{++} & \mbox{\boldmath $\vec{\sigma}$}^{(1)}\cdot  
\mbox{\boldmath
$\vec{p}$}\psi _{+-}
\end{array}
\right)
-m_1\left(
\begin{array}{cc}
\psi _{++} & \psi _{+-} \\
-\psi _{-+} & -\psi _{--}
\end{array}
\right)+ \nonumber \\
&+&\left(
\begin{array}{cc}
\mbox{\boldmath $\vec{\sigma}$}^{(2)}\cdot
\mbox{\boldmath  
$\vec{p}$}\psi _{+-} & \mbox{\boldmath $\vec{\sigma}$}^{(2)}\cdot
\mbox{\boldmath
$\vec{p}$}\psi _{++} \\
\mbox{\boldmath $\vec{\sigma}$}^{(2)}\cdot
\mbox{\boldmath
$\vec{p}$}\psi _{--} & \mbox{\boldmath $\vec{\sigma}$}^{(2)}\cdot
\mbox{\boldmath
$\vec{p}$}\psi _{-+}
\end{array}
\right)-
m_2\left(
\begin{array}{cc}
\psi _{++} & -\psi _{+-} \\
\psi _{-+} & -\psi _{--}
\end{array}\right)+
S(r)\left( 
\begin{array}{cc}
\psi _{++} & -\psi _{+-} \\
-\psi _{-+} & \psi _{--}
\end{array}
\right)- \nonumber \\
&-&P(r)\left( 
\begin{array}{cc}
\psi _{--} & -\psi _{-+} \\
-\psi _{+-} & \psi _{++}
\end{array}
\right)
-V(r)\left(
\begin{array}{cc}
\psi _{++} & \psi _{+-} \\
\psi _{-+} & \psi _{--} 
\end{array}
\right)=0
\label{equat:ion1}
\end{eqnarray}
In order to simplify Eq.(\ref{equat:ion1}) and bring it to a form which can be handled
easily, we introduce the following components \cite{a17,a18} 
\begin{eqnarray}
\begin{array}{ccl}
\left. \begin{array}{ll}\phi \\
        \phi^0
        \end{array} \right \} &=& 
P_0\frac{i}{\sqrt{2}}(\psi_{++}\mp\psi_{--}) \\
\left. \begin{array}{ll}\boldmath \vec{\phi} \\
        \boldmath \vec{\phi^0}
        \end{array} \right \} &=& 
\frac{1}{2}(\mbox{\boldmath $ \vec{\sigma}$}^{(1)}-\mbox{\boldmath $
\vec{\sigma}$}^{(2)})P_1\frac{1}{\sqrt{2}}(\psi_{+-}\pm\psi_{-+}) \\
\end{array}      
\label{fun:ction1}
\end{eqnarray}
\begin{eqnarray}
\begin{array}{ccl}
\left.\begin{array}{ll}\chi \\
        \chi^0
        \end{array} \right \} &=& 
P_0\frac{i}{\sqrt{2}}(\psi_{+-}\mp\psi_{-+})  \\
\left. \begin{array}{ll}\boldmath \vec{\chi} \\
        \boldmath \vec{\chi^0}
        \end{array} \right \} &=& 
\frac{1}{2}(\mbox{\boldmath $\vec{\sigma}$}^{(1)}-
\mbox{\boldmath $\vec{\sigma}$}^{(2)})P_1\frac{1}{\sqrt{2}}(\psi_
{++}\pm\psi_{--}) 
\end{array}
\label{fun:ction2}
\end{eqnarray} 
where
\begin{eqnarray}
\begin{array}{ccl}
P_0 &=& \frac 14\left(1-\mbox{\boldmath $\vec{\sigma}$}^{(1)}\cdot
\mbox{\boldmath $\vec{\sigma}$}^{(2)}\right)\\
P_1 &=& \frac 14\left(3+
\mbox{\boldmath $\vec{\sigma}$}^{(1)}\cdot \mbox{\boldmath
$\vec{\sigma}$}^{(2)}\right)
\end{array} 
\end{eqnarray}
 are the projection operators on states with total
spin $S=0$ and $S=1$, respectively, 
\begin{eqnarray}
\begin{array}{lcl}
P_0 |\rm state\rangle &=& 
|\rm state \mbox{ } S=0\rangle \\
P_1 |\rm state\rangle &=&
|\rm state \mbox{ } S=1\rangle\end{array}
\end{eqnarray}
 The components $\phi ,\phi ^0,\chi ,\chi ^0$ and $\boldmath
\vec \phi ,\mbox{\boldmath $ \vec{\phi^0}$},\boldmath\vec \chi ,
\mbox{\boldmath $
\vec{\chi^0}$}$ correspond to spin $S=0$ and $S=1$, respectively and they are
functions of \mbox{\boldmath $\vec r$}. Noting that  
\begin{eqnarray}
\begin{array}{ccl}
P_0(\mbox{\boldmath $\vec{\sigma}$}^{(1)}-\mbox{\boldmath
$\vec{\sigma}$}^{(2)}) &=& 
(\mbox{\boldmath $\vec{\sigma}$}^{(1)}-\mbox{\boldmath
$\vec{\sigma}$}^{(2)})P_1 \\
P_0(\mbox{\boldmath $\vec{\sigma}$}^{(1)}+\mbox{\boldmath
$\vec{\sigma}$}^{(2)}) &=& 
(\mbox{\boldmath $\vec{\sigma}$}^{(1)}+\mbox{\boldmath
$\vec{\sigma}$}^{(2)})P_0=0
\end{array}
\label{ek:fra}
\end{eqnarray}
and by means of the identities 
\begin{eqnarray}
\begin{array}{ccl}
P_0(\sigma _i^{(1)}\sigma _k^{(2)}+\sigma _k^{(1)}\sigma _i^{(2)}) & = & 
-2\delta _{ik}P_0 \\ 
P_0(\sigma _i^{(1)}\sigma _k^{(2)}-\sigma _k^{(1)}\sigma _i^{(2)}) & = & 
iP_0\epsilon _{ikl}(
\mbox{\boldmath $\vec{\sigma}$}^{(1)}-\mbox{\boldmath
$\vec{\sigma}$}^{(2)})_l \\ \hookrightarrow P_0(\sigma _i^{(1)}-\sigma
_i^{(2)})(\sigma _k^{(1)}\pm \sigma _k^{(2)}) & = & \left\{ 
\begin{array}{l}
2iP_0\epsilon _{ikl}(
\mbox{\boldmath $\vec{\sigma}$}^{(1)}-\mbox{\boldmath
$\vec{\sigma}$}^{(2)})_l \\ 2P_0\delta _{ik}
\end{array}
\right. 
\end{array}
\end{eqnarray}
Eq.(\ref{equat:ion1}) leads to the following set of component wave equations \cite{a10,a17,a18}
\begin{eqnarray}
\begin{array}{lcc}
\frac{1}{2}\biggl[E+S(r)-P(r)-V(r)\biggr]
\phi^0-\frac{(m_1+m_2)}{2}\phi-i\mbox{\boldmath
$\vec{p}$}\cdot \mbox{\boldmath $\vec{\phi}$} & = & 0\\
\frac{1}{2}\biggl[E+S(r)+P(r)-V(r)\biggr]\phi-\frac{(m_1+m_2)}{2}\phi^0 &=& 0 \\
\frac{1}{2}\biggl[E-S(r)+P(r)-V(r)\biggr]\chi^0-\frac{(m_1-m_2)}{2}\chi-i\mbox{\boldmath
$\vec{p}$}\cdot \mbox{\boldmath  
$\vec{\chi}$} &=& 0 \\
\frac{1}{2}\biggl[E-S(r)-P(r)-V(r)\biggr]\chi-\frac{(m_1-m_2)}{2}\chi^0 &=& 0 \\
\frac{1}{2}\biggl[E+S(r)-P(r)-V(r)\biggr]\mbox{\boldmath $\vec{\chi}$}-\frac{(m_1+m_2)}{2}
\mbox{\boldmath   
$\vec{\chi^0}$}+i\mbox{\boldmath $\vec{p}$}\chi^0 &=& 0 \\
\frac{1}{2}\biggl[E+S(r)+P(r)-V(r)\biggr]\mbox{\boldmath
$\vec{\chi^0}$}-\frac{(m_1+m_2)}{2}
\mbox{\boldmath
$\vec{\chi}$}+i\mbox{\boldmath $\vec{p}$}\times\mbox{\boldmath
$\vec{\phi^0}$} &=& 0 \\
\frac{1}{2}\biggl[E-S(r)+P(r)-V(r)\biggr]\mbox{\boldmath
$\vec{\phi}$}-\frac{(m_1-m_2)}{2}\mbox{\boldmath $\vec{\phi}^0$}+
i\mbox{\boldmath
$\vec{p}$}\phi^0 &=& 0 \\
\frac{1}{2}\biggl[E-S(r)-P(r)-V(r)\biggr]\mbox{\boldmath
$\vec{\phi}^0$}-\frac{(m_1-m_2)}{2}\mbox{\boldmath
$\vec{\phi}$}+i\mbox{\boldmath
$\vec{p}$}\times\mbox{\boldmath $\vec{\chi}^0$} &=& 0 
\end{array}
\label{eq:bre}
\end{eqnarray}
The type of the potential we consider is central therefore the next 
step we will follow is the introduction of a method that will 
eventually separate
the angular from the radial dependences as it was performed in the three
dimensional Schr\"odinger equation. To this end, it is convenient to
introduce the derivative 
\begin{equation}
\frac \partial {\partial n_{i\perp }}\equiv (\delta _{ik}-n_in_k)\frac
\partial {\partial n_k}
\end{equation}
where $\mbox{\boldmath$\vec{n}$}=\frac{\mbox{\boldmath$\vec{r}$}}r$ is a
three dimensional vector, while
$\frac \partial {\partial \mbox{\boldmath $\vec{n}$}_{\perp}}$ 
lies in the perpendicular plane $\left[%
n_i\frac \partial {\partial n_{i{\perp }}}=n_i(\delta _{ik}-n_in_k)\frac
\partial {\partial n_k}=n_k\frac \partial {\partial n_k}-n_k\frac \partial
{\partial n_k}=0\right]$. Some useful identities that are going to be used are the
following 
\begin{eqnarray}
\begin{array}{lcl}
\frac{\partial}{\partial
\mbox{\boldmath$\vec{r}$}} &=& \frac{1}{r}\frac{\partial}{\partial
\mbox{\boldmath$\vec{n}$}_{\perp}}+
\mbox{\boldmath$\vec{n}$}\frac{\partial}{\partial
 r} \\
\frac{\partial}{\partial 
\mbox{\boldmath$\vec{n}$}_{\perp}}\cdot
\mbox{\boldmath$\vec{n}$} &=& (\delta_{ik}-n_in_k)\frac{\partial}
{\partial n_{k}}n_i=
(\delta_{ik}-n_in_k)\delta_{ki}=3-1=2
\end{array}
\end{eqnarray}
Then, $\mbox{\boldmath$\vec{\bigtriangledown}$}^2=\frac{\partial ^2}{%
\partial \mbox{\boldmath$\vec{r}$}^2}=(\frac 1r\frac \partial {\partial 
\mbox{\boldmath$\vec{n}$}_{\perp}}+\mbox{\boldmath$\vec{n}$}\frac \partial
{\partial r})^2=\frac{\partial ^2}{\partial r^2}+\frac 1{r^2}\frac{\partial
^2}{\partial \mbox{\boldmath$\vec{n}$}_{\perp}^2}+\frac 2r\frac \partial
{\partial r}$. This form of the Laplacian indicates that $\frac{\partial ^2}{%
\partial \mbox{\boldmath$\vec{n}$}_{\perp}^2}$ was correctly regarded as 
the angular part of the Laplacian.
\subsection{Expansion of the wavefunctions}

The component functions defined by (\ref{fun:ction1}) and 
(\ref{fun:ction2}) are classified in two groups: the scalar 
($\phi ,\phi ^0,\chi
,\chi ^0$) which refer to states with $S=0$, 
and the vector ($\boldmath
\vec \phi ,\mbox{\boldmath $ \vec{\phi^0}$},\boldmath\vec \chi ,%
\mbox{\boldmath $
\vec{\chi^0}$})$) which correspond to states with $S=1$ \cite{a10,a17,a18,a20}. These functions 
can be expanded in the following way:\\ \underline{(i). Scalar
functions ($S=0$)}.\\ The scalar functions can be written as 
\begin{equation}
B(\mbox{\boldmath $\vec{r}$})=\sum_{j,m}B(r)Y_{jm}(\mbox{\boldmath $\vec{n}$})
\label{sca:fun}
\end{equation}
where $B(\mbox{\boldmath $\vec{r}$})$ stands for any of the four $S=0$
component functions, $B(r)$ is the radial part of the function and $Y_{jm}(%
\mbox{\boldmath $\vec{n}$})$ are the spherical harmonics depending only on
the angles. \\ \underline{(ii). Vector functions ($S=1$)}.\\ We define an
operator $S_k$ which acts on any of the four vector components 
[(\ref{fun:ction1}), (\ref{fun:ction2})] as
follows 
\begin{equation}
S_k\mbox{(\boldmath $\vec{A}$}(\mbox{\boldmath
$\vec{r}$}))|_i\equiv \frac 12(\mbox{\boldmath $
\vec{\sigma}$}^{(1)}-\mbox{\boldmath $
\vec{\sigma}$}^{(2)})_iP_1\left(\frac 12(\mbox{\boldmath $
\vec{\sigma}$}^{(1)}+\mbox{\boldmath $
\vec{\sigma}$}^{(2)})_k)\right)\frac 1{\sqrt{2}}(...)
\end{equation}
where $(...)$ equals $(\psi _{+-}+\psi _{-+})$ (for 
\mbox{\boldmath $ \vec{\phi}$}), 
$(\psi _{+-}-\psi _{-+})$ (for \mbox{\boldmath $ \vec{\phi^0}$} ), 
$(\psi _{++}+\psi
_{--})$ (for \mbox{\boldmath $ \vec{\chi}$}), 
$(\psi _{++}-\psi _{--})$ (for \mbox{\boldmath $ \vec{\chi^0}$}) 
and $\mbox{\boldmath $\vec{A}$}(%
\mbox{\boldmath
$\vec{r}$})$ is any of the four vector components. We notice that the
operator $S_k$ does not act on \mbox{\boldmath
$\vec{A}$}(\mbox{\boldmath
$\vec{r}$}) from the left since in that case it would give zero $\left(
\frac 12(\mbox{\boldmath $
\vec{\sigma}$}^{(1)}+\mbox{\boldmath $
\vec{\sigma}$}^{(2)})_k\frac 12(\mbox{\boldmath $
\vec{\sigma}$}^{(1)}-\mbox{\boldmath $
\vec{\sigma}$}^{(2)})_iP_1\stackrel{(\ref{ek:fra})}{=}0\right) $. 
From the definition
of $S_k\mbox{(\boldmath $\vec{A}$}(\mbox{\boldmath
$\vec{r}$}))|_i$, it is clear that $S_k\mbox{(\boldmath $\vec{A}$}(%
\mbox{\boldmath
$\vec{r}$}))|_i=i\epsilon _{ikl}$ $\phi _l$. By making use
of the multipole technique \cite{a10,a17,a18}, 
it is possible to expand any vector component $%
\mbox{\boldmath $\vec{A}$}(\mbox{\boldmath
$\vec{r}$})$ into three parts: ``electric'' ($A_e$), 
``longitudinal'' ($A_l$) and
``magnetic'' ($A_m$)defined as 
$$
\begin{array}{ccc}
A_e(\mbox{\boldmath $\vec{r}$}) & = & {\displaystyle \sum_{j,m}}A_e(r)Y_{jm}( 
\mbox{\boldmath $\vec{n}$}) \\ A_l(\mbox{\boldmath $\vec{r}$}) & = & 
{\displaystyle \sum_{j,m}}A_l(r)Y_{jm}( 
\mbox{\boldmath $\vec{n}$}) \\ A_m(\mbox{\boldmath $\vec{r}$}) & = & 
{\displaystyle \sum_{j,m}}A_m(r)Y_{jm}(\mbox{\boldmath $\vec{n}$})\label{c:l} 
\end{array}
$$
\begin{equation}
\label{d:m}\mbox{\boldmath $\vec{A}$}(\mbox{\boldmath
$\vec{r}$})=\underbrace{\mbox{\boldmath
$\vec{n}$}A_e(\mbox{\boldmath $\vec{r}$})}_{\mbox{\boldmath $\vec{A}$}_e(
\mbox{\boldmath
$\vec{r}$})}-
\underbrace{\frac \partial {\partial \mbox{\boldmath
$\vec{n}$}_{\perp }}\frac{A_l(\mbox
{\boldmath $\vec{r}$})}{j(j+1)}}_{\mbox{\boldmath $\vec{A}$}_l(\mbox{\boldmath
$\vec{r}$})}
-\underbrace{(\mbox{\boldmath
$\vec{n}$}\times \frac \partial {\partial \mbox{\boldmath
$\vec{n}$}_{\perp }})\frac{A_m(\mbox{\boldmath $\vec{r}$})}{j(j+1)}}_
{\mbox{\boldmath $\vec{A}$}_m(%
\mbox{\boldmath
$\vec{r}$})} 
\label{vec:fun}
\end{equation}
(In the rest of our study, for the sake of simplicity, we will the drop
the summation symbol ${\displaystyle \sum_{j,m}}$). 
It may seem that $j$ in $Y_{jm}$ is the orbital momentum
rather than the total one. But this is not true for the function $%
\mbox{\boldmath $\vec{A}$}(\mbox{\boldmath
$\vec{r}$})$ because the vectors to which it is proportional, depend on
angles. Actually, if $\mbox{\boldmath $\vec{S}$}$ and $%
\mbox{\boldmath
$\vec{L}$}$ are the total spin and total orbital angular momentum operators, 
respectively, then for each part of $\mbox{\boldmath $\vec{A}$}(%
\mbox{\boldmath
$\vec{r}$})$, for instance $\mbox{\boldmath $\vec{A}$}_e(%
\mbox{\boldmath
$\vec{r}$})\equiv \mbox{\boldmath
$\vec{n}$}A_e(r)Y_{jm}(\mbox{\boldmath
$\vec{n}$})$, one has \cite{a10}
\begin{eqnarray*}
J_k\mbox{(\boldmath $\vec{A}$}_e(\mbox{\boldmath
$\vec{r}$}))|_i &=&S_k\mbox{(\boldmath $\vec{A}$}_e(\mbox{\boldmath
$\vec{r}$}))|_i+L_k\mbox{(\boldmath $\vec{A}$}_e(\mbox{\boldmath
$\vec{r}$}))|_i=\\
&=&i\epsilon _{ikl}n_lA_e(r)Y_{jm}(\mbox{\boldmath
$\vec{r}$})-
i\epsilon _{kla}n_l\frac \partial {\partial n_a}(n_iA_e(r)Y_{jm}(%
\mbox{\boldmath
$\vec{n}$}))=\\
&=&-i\epsilon _{kla}n_ln_iA_e(r)\frac{\partial Y_{jm}(\mbox{\boldmath
$\vec{n}$})}{\partial n_a}=\mbox{\boldmath
$\vec{n}_i$}L_k(A_e(r)Y_{jm}(\mbox{\boldmath
$\vec{n}$}))\Longrightarrow \\
& \Longrightarrow &\mbox{\boldmath
$\vec{J}$}^2\mbox{\boldmath $\vec{A}$}_e(\mbox{\boldmath
$\vec{r}$})=\mbox{\boldmath
$\vec{n}$}L^2(A_e(r)Y_{jm}(\mbox{\boldmath
$\vec{n}$}))=j(j+1)\mbox{\boldmath $\vec{A}$}_e(\mbox{\boldmath
$\vec{r}$}) 
\end{eqnarray*}
which implies that $j$ is the total angular momentum.
By inserting [(\ref{sca:fun}),(\ref{vec:fun})] into Eqs.(\ref{eq:bre}), 
we manage to eliminate the angular
dependences and the Breit equation reduces to the following set of
sixteen radial equations  \\
\begin{eqnarray}
\begin{array}{lcc}
\frac{1}{2}\biggl[E+S(r)-P(r)-V(r)\biggr]\phi^0-\frac{(m_1+m_2)}{2}\phi-
(\frac{d}{dr}+\frac{2}{r})\phi_e-\frac{1}{r}\phi_l &=& 0 \\
\frac{1}{2}\biggl[E+S(r)+P(r)-V(r)\biggr] \phi-\frac{(m_1+m_2)}{2}\phi^0 &=& 0 \\
\frac{1}{2}\biggl[E-S(r)+P(r)-V(r)\biggr]\chi^0-\frac{(m_1-m_2)}{2}\chi-(\frac{d}{dr}+
\frac{2}{r})\chi_e-\frac{1}{r}\chi_l &=& 0 \\
\frac{1}{2}\biggl[E-S(r)-P(r)-V(r)\biggr]\chi-\frac{(m_1-m_2)}{2}\chi^0 &=& 0 \\
\frac{1}{2}\biggl[E+S(r)-P(r)-V(r)\biggr]\chi_e-\frac{(m_1+m_2)}{2}\chi^0_e+
\frac{d}{dr}\chi^0 &=& 0 \\
\frac{1}{2}\biggl[E+S(r)-P(r)-V(r)\biggr]\chi_l-
\frac{(m_1+m_2)}{2}\chi^0_l-\frac{j(j+1)}{r}\chi^0 &=& 0 \\
\frac{1}{2}\biggl[E+S(r)-P(r)-V(r)\biggr]\chi_m-\frac{(m_1+m_2)}{2}\chi^0_m &=& 0 \\
\frac{1}{2}\biggl[E+S(r)+P(r)-V(r)\biggr]\chi^0_e-\frac{(m_1+m_2)}{2}\chi_e+
\frac{1}{r}\phi^0_m &=& 0 \\
\frac{1}{2}\biggl[E+S(r)+P(r)-V(r)\biggr]\chi^0_l-\frac{(m_1+m_2)}{2}\chi_l
-(\frac{d}{dr}+\frac{1}{r})\phi^0_m &=& 0 \\
\frac{1}{2}\biggl[E+S(r)+P(r)-V(r)\biggr]\chi^0_m-\frac{(m_1+m_2)}{2}\chi_m+
\frac{j(j+1)}{r}\phi^0_e+(\frac{d}{dr}+\frac{1}{r})\phi^0_l  &=& 0 \\
\frac{1}{2}\biggl[E-S(r)+P(r)-V(r)\biggr]\phi_e-\frac{(m_1-m_2)}{2}\phi^0_e+
\frac{d}{dr}\phi^0 &=& 0 \\
\frac{1}{2}\biggl[E-S(r)+P(r)-V(r)\biggr]\phi_l-\frac{(m_1-m_2)}{2}\phi^0_l-
\frac{j(j+1)}{r}\phi^0  &=& 0 \\
\frac{1}{2}\biggl[E-S(r)+P(r)-V(r)\biggr]\phi_m-\frac{(m_1-m_2)}{2}\phi^0_m &=& 0 \\
\frac{1}{2}\biggl[E-S(r)-P(r)-V(r)\biggr]\phi^0_e-\frac{(m_1-m_2)}{2}\phi_e+
\frac{1}{r}\chi^0_m &=& 0 \\
\frac{1}{2}\biggl[E-S(r)-P(r)-V(r)\biggr]\phi^0_l-\frac{(m_1-m_2)}{2}\phi_l-
(\frac{d}{dr}+\frac{1}{r})\chi^0_m &=& 0 \\
\frac{1}{2}\biggl[E-S(r)-P(r)-V(r)\biggr]\phi^0_m-\frac{(m_1-m_2)}{2}\phi_m+
\frac{j(j+1)}{r}\chi^0_e+(\frac{d}{dr}+\frac{1}{r})\chi^0_l &=& 0
\end{array}
\label{set:equ}
\end{eqnarray}

At this point, we will concentrate on the properties of the scalar and
vector components as well as the new components to which they have been
expanded. As pointed out earlier, the scalar components $\phi ,\phi ^0,\chi
,\chi ^0$ describe the $S=0$ states, while the rest, the vector
components, describe 
the states with $S=1.$ In the latter case, for 
$\mbox{\boldmath $\vec{\chi}$}$, there are three states
characterised by the spectroscopic signatures $^3(j-1)_j,^3(j+1)_j$ and $%
^3j_j$ (in the atomic notation $^{2S+1}L_j$) described by the wave-functions 
$\chi _{l=j-1},\chi _{l=j+1}$ and $\chi _l$ satisfying the following
relations \cite{a17,a18} 
\begin{eqnarray}
\begin{array}{ccl}
\chi _e & = & \sqrt{\frac j{2j+1}}\chi _{l=j-1}+\sqrt{\frac{j+1}{2j+1}}\chi
_{l=j+1} \\ \chi _l & = & \sqrt{j(j+1)}(-\sqrt{\frac{j+1}{2j+1}}\chi
_{l=j-1}+\sqrt{\frac j{2j+1}}\chi _{l=j+1}) \\ \chi _m & = & \sqrt{j(j+1)}%
\chi _{l=j}
\label{el:m}
\end{array}
\end{eqnarray}
Certainly, this can be generalised for the rest of the $S=1$ components: $%
\mbox{\boldmath $\vec{\chi}$}^0,\mbox{\boldmath $\vec{\phi}$}$ and $%
\mbox{\boldmath $\vec{\phi}$}^0$. The ``magnetic'' component is chosen to
have $l=j$ while the ``electric'' and the ``longitudinal'' components have $%
l=j\pm 1$ mixed if $j>0$ (and $l=1$, if $j=0$). If $j=0$, then there should be
only one $\chi $ component and we note this is satisfied since $\chi _l=\chi
_m=0$ and $\chi _{l=j-1}=\chi _{l=j}=0,\chi _{l=j+1}=\chi _e$. There are
only two states with $j=0$, $^1S_0$ ($j=0,l=0,S=0$) and $^3P_0$ ($%
j=0,l=1,S=1 $).\\ 
The expressions (\ref{fun:ction1}) and 
(\ref{fun:ction2}) indicate that, for $S=0$, the
``large-large'' ($\psi _{++}$) and ``small-small'' ($\psi _{--}$) components
are contained in the $\mbox{\boldmath $\vec{\chi}$},%
\mbox{\boldmath $\vec{\chi}$}^0$ and the ``large-small'' ($\psi _{+-}$) and
``small-large'' ($\psi _{-+}$) components appear in the $%
\mbox{\boldmath $\vec{\phi}$},\mbox{\boldmath
$\vec{\phi}$}^0$. In addition, all ``electric'' and ``longitudinal''
components of the $\mbox{\boldmath $\vec{\chi}$},%
\mbox{\boldmath $\vec{\chi}$}^0,\mbox{\boldmath $\vec{\phi}$},%
\mbox{\boldmath $\vec{\phi}$}^0$ correspond to states with magnetic quantum
number $m_s=\pm 1$ while the ``magnetic'' components describe states with $%
m_s=0$. On the other hand, the states with $S=0$ have the spectroscopic
notation $^1j_j$ and the ``large-large'' ($\psi _{++}$) and ``small-small'' (%
$\psi _{--}$) components are contained in the $\phi ,\phi ^0$ and the
``large-small'' ($\psi _{+-}$) and ``small-large'' ($\psi _{-+}$) components
appear in the $\chi ,\chi ^0$. All $\phi ,\phi ^0,\chi ,\chi ^0$ correspond
to states with $m_s=0$.

The parity of the system equals $P=\eta (-1)^l$, where $\eta =1$ or $-1$ for
fermion-fermion or fermion-antifermion system, respectively. The sixteen
components have the following parity: \\(i). $\phi ,\phi ^0:P=\eta (-1)^j$
because they both describe the $S=0$ states, therefore $l=j$,\\ (ii). $\phi
_e,\phi _e^0,\phi _l,\phi _l^0:P=-\eta (-1)^{j+1}$ (or $-\eta (-1)^{j-1}$
which gives the same result) $=\eta (-1)^j$. According to (\ref{el:m}) the
``electric'' and ``longitudinal'' components are combinations of states with 
$l=j+1$ and $l=j-1$ and this justifies the exponent $j+1$ (or $j-1$). The $%
(-)$ sign in front of $\eta $ is due to the fact that all these functions
are combinations of ``small'' and ``large'' components,\\ (iii). $\chi
_m,\chi _m^0:P=\eta (-1)^j$ ((\ref{el:m}) implies that the ``magnetic''
component corresponds to $l=j$ states),\\ (iv). $\chi ,\chi ^0:P=-\eta (-1)^j$%
. Both $\chi ,\chi ^0$ describe $S=0$ states therefore $l=j$, however, in
contrast to the parity of the $\phi ,\phi ^0$, there is a minus sign which
is attributed to the fact that $\chi ,\chi ^0$ are combinations of ``small''
and ``large'' components, \\ (v). $\chi _e,\chi _e^0,\chi _l,\chi
_l^0:P=-\eta (-1)^j$, (as in (ii)),\\ (vi). $\phi _m,\phi _m^0:P=-\eta
(-1)^j,$ (as in (iii)).\\Again, the combination of ``small'' and ``large''
components accounts for the minus sign in (v), (vi). To summarise, the
components $\phi ,\phi ^0,\phi _e,\phi _e^0,\phi _l,\phi _l^0,\chi _m,\chi
_m^0$ have parity $P=\eta (-1)^j$ while the components $\chi ,\chi ^0,\chi
_e,\chi _e^0,\chi _l,\chi _l^0,\phi _m,\phi _m^0$ have parity $P=-\eta
(-1)^j $. We call the former case, Pseudoscalar Particle Trajectory (PPT)
while the latter is called Vector Particle Trajectory (VPT) \cite{a17,a18}.
The ``large-large'' and ``small-small'' components in the PPT are contained
in the $\phi ,\phi ^0,\chi _m,\chi _m^0$ and have spectroscopic signature $%
^1j_j $ or $^3j_j.$ In the VPT, the ``large-large'' and ``small-small''
components are contained in $\chi _e,\chi _e^0,\chi _l,\chi _l^0$ and have
signature $^3(j-1)_j$ or $^3(j+1)_j.$

Returning to Eqs.(\ref{set:equ}) and by making use of the previous discussion,
that set of equations can be split into two sets of equations according to
whether they belong to the PPT or VPT regime. To be more specific, if the two
fermions do not possess the same mass, the following two sets of equations are
obtained from Eqs.(\ref{set:equ}) \cite{a10,a17,a18}

\underline{(i). PPT, $^1j_j\mbox{  } {\rm or} \mbox{  }3j_j,
P=\eta (-1)^j$.} 
\begin{eqnarray}
\begin{array}{lcc}
\frac{1}{2}\biggl[E+S(r)-P(r)-V(r)\biggr]\phi^0-\frac{(m_1+m_2)}{2}\phi-
(\frac{d}{dr}+\frac{2}{r})\phi_e-\frac{1}{r}\phi_l &=& 0 \\
\frac{1}{2}\biggl[E+S(r)+P(r)-V(r)\biggr]\phi-\frac{(m_1+m_2)}{2}\phi^0 &=& 0 \\
\frac{1}{2}\biggl[E+S(r)-P(r)-V(r)\biggr]\chi_m-\frac{(m_1+m_2)}{2}\chi^0_m &=& 0 \\
\frac{1}{2}\biggl[E+S(r)+P(r)-V(r)\biggr]\chi^0_m-\frac{(m_1+m_2)}{2}\chi_m+
\frac{j(j+1)}{r}\phi^0_e+(\frac{d}{dr}+\frac{1}{r})\phi^0_l  &=& 0 \\
\frac{1}{2}\biggl[E-S(r)+P(r)-V(r)\biggr]\phi_e-\frac{(m_1-m_2)}{2}\phi^0_e+
\frac{d}{dr}\phi^0 &=& 0 \\
\frac{1}{2}\biggl[E-S(r)+P(r)-V(r)\biggr]\phi_l-\frac{(m_1-m_2)}{2}\phi^0_l-
\frac{j(j+1)}{r}\phi^0  &=& 0 \\
\frac{1}{2}\biggl[E-S(r)+P(r)-V(r)\biggr]\phi_m-\frac{(m_1-m_2)}{2}\phi^0_m &=& 0 \\
\frac{1}{2}\biggl[E-S(r)-P(r)-V(r)\biggr]\phi^0_e-\frac{(m_1-m_2)}{2}\phi_e+
\frac{1}{r}\chi^0_m &=& 0 \\
\frac{1}{2}\biggl[E-S(r)-P(r)-V(r)\biggr]\phi^0_l-\frac{(m_1-m_2)}{2}\phi_l-
(\frac{d}{dr}+\frac{1}{r})\chi^0_m &=& 0 
\end{array}
\end{eqnarray}
\underline{(ii). VPT, $^3(j\pm 1)_j,P=-\eta (-1)^j$.} 
\begin{eqnarray}
\begin{array}{lcc}
\frac{1}{2}\biggl[E-S(r)+P(r)-V(r)\biggr]\chi^0-\frac{(m_1-m_2)}{2}\chi-(\frac{d}{dr}+
\frac{2}{r})\chi_e-\frac{1}{r}\chi_l &=& 0 \\
\frac{1}{2}\biggl[E-S(r)-P(r)-V(r)\biggr]\chi-\frac{(m_1-m_2)}{2}\chi^0 &=& 0 \\
\frac{1}{2}\biggl[E+S(r)-P(r)-V(r)\biggr]\chi_e-\frac{(m_1+m_2)}{2}\chi^0_e+
\frac{d}{dr}\chi^0 &=& 0 \\
\frac{1}{2}\biggl[E+S(r)-P(r)-V(r)\biggr]\chi_l-
\frac{(m_1+m_2)}{2}\chi^0_l-\frac{j(j+1)}{r}\chi^0 &=& 0 \\
\frac{1}{2}\biggl[E+S(r)+P(r)-V(r)\biggr]\chi^0_e-\frac{(m_1+m_2)}{2}\chi_e+
\frac{1}{r}\phi^0_m &=& 0 \\
\frac{1}{2}\biggl[E+S(r)+P(r)-V(r)\biggr]\chi^0_l-\frac{(m_1+m_2)}{2}\chi_l
-(\frac{d}{dr}+\frac{1}{r})\phi^0_m &=& 0 \\
\frac{1}{2}\biggl[E-S(r)-P(r)-V(r)\biggr]\phi^0_m-\frac{(m_1-m_2)}{2}\phi_m+
\frac{j(j+1)}{r}\chi^0_e+(\frac{d}{dr}+\frac{1}{r})\chi^0_l &=& 0
\end{array}
\end{eqnarray}

On the other hand, in the case of equal masses $m_1=m_2=m$, the set PPT
splits into two subsets, one with spectroscopic signature $^1j_j$ and
a second with $^3j_j$ while the VPT remains 
unchanged\\ \underline{(i). PPT, 
$^1j_j,P=\eta (-1)^j$.} 
\begin{eqnarray}
\begin{array}{lcc}
\frac{1}{2}\biggl[E+S(r)-P(r)-V(r)\biggr]\phi^0-m\phi-
(\frac{d}{dr}+\frac{2}{r})\phi_e-\frac{1}{r}\phi_l &=& 0 \\
\frac{1}{2}\biggl[E+S(r)+P(r)-V(r)\biggr]\phi-m\phi^0 &=& 0 \\
\frac{1}{2}\biggl[E-S(r)+P(r)-V(r)\biggr]\phi_e+
\frac{d}{dr}\phi^0 &=& 0 \\
\frac{1}{2}\biggl[E-S(r)+P(r)-V(r)\biggr]\phi_l-
\frac{j(j+1)}{r}\phi^0  &=& 0  
\label{b:a}
\end{array}
\end{eqnarray}
\underline{(ii). PPT, $^3j_j,P=\eta (-1)^j$ .} 
\begin{eqnarray}
\begin{array}{lcc}
\frac{1}{2}\biggl[E+S(r)-P(r)-V(r)\biggr]\chi_m-m\chi^0_m &=& 0 \\
\frac{1}{2}\biggl[E+S(r)+P(r)-V(r)\biggr]\chi^0_m-m\chi_m+
\frac{j(j+1)}{r}\phi^0_e+(\frac{d}{dr}+\frac{1}{r})\phi^0_l  &=& 0 \\
\frac{1}{2}\biggl[E-S(r)-P(r)-V(r)\biggr]\phi^0_e+
\frac{1}{r}\chi^0_m &=& 0 \\
\frac{1}{2}\biggl[E-S(r)-P(r)-V(r)\biggr]\phi^0_l-
(\frac{d}{dr}+\frac{1}{r})\chi^0_m &=& 0
\label{b:c}
\end{array}
\end{eqnarray}
\underline{(iii). VPT, $^3(j\pm 1)_j,P=-\eta (-1)^j$ .} 
\begin{eqnarray}
\begin{array}{lcc}
\frac{1}{2}\biggl[E-S(r)+P(r)-V(r)\biggr]\chi^0-(\frac{d}{dr}+
\frac{2}{r})\chi_e-\frac{1}{r}\chi_l &=& 0 \\
\frac{1}{2}\biggl[E+S(r)-P(r)-V(r)\biggr]\chi_e-m\chi^0_e+
\frac{d}{dr}\chi^0 &=& 0 \\
\frac{1}{2}\biggl[E+S(r)-P(r)-V(r)\biggr]\chi_l-
m\chi^0_l-\frac{j(j+1)}{r}\chi^0 &=& 0 \\
\frac{1}{2}\biggl[E+S(r)+P(r)-V(r)\biggr]\chi^0_e-m\chi_e+
\frac{1}{r}\phi^0_m &=& 0 \\
\frac{1}{2}\biggl[E+S(r)+P(r)-V(r)\biggr]\chi^0_l-m\chi_l
-(\frac{d}{dr}+\frac{1}{r})\phi^0_m &=& 0 \\
\frac{1}{2}\biggl[E-S(r)-P(r)-V(r)\biggr]\phi^0_m+
\frac{j(j+1)}{r}\chi^0_e+(\frac{d}{dr}+\frac{1}{r})\chi^0_l &=& 0
\label{b:d}
\end{array}
\end{eqnarray}
The remaining two components $\chi ,\phi _m$ vanish in the event of the
particles having the same mass $m$.


\section{A Potential for quarkonium}
\label{sec:three}
In the Introduction, we discussed the lack of covariance
characterising the Breit equation. The wavefunction of the system of the
two fermions depends on the position of the two particles (that is, two
vectors, each for the two particles, are necessary) while it depends 
only on one time variable.
This fact does not allow a Lorentz-invariant formulation. However, if a
very strong, short-range potential is considered to govern the interaction
of the two particles, we can assume the two particles are so close
to one another that just one position vector is
enough to describe the motion and the quantum mechanics of the system.
In this case, the number of the position vector variables equals that of
the time variables and we can assume that the equation becomes 
approximately covariant. The question rising is whether there exist such
interactions between particles and if there are systems of fermions to
which the Breit equation can be applied and subsequently be tested. 
Talking about small distances, we are led to the plausible question
of what happens at distances up to 1 fermi which is a feature 
of the strong interactions between quarks. It is known
that there are systems, quarkonia, which are bound states of
quarks and antiquarks. Actually, 
experimental data about many of these states exist \cite{a35},
therefore, one could maintain that they constitute a laboratory where the
validity of any theory aiming at describing them can be tested. 

In the non-relativistic limit, the quark-antiquark interaction is 
dominated by vector and scalar potential \cite{a10,b34}. The former, which is 
of the Coulomb type 
consitutes the interaction at small distances and corresponds to 
one gluon exchange. In that case, 
quarkonium can be viewed as the ``hydrogen atom'' of strong 
interaction Physics. By contrast, at large distances, confinement 
dominates and the interaction is proportional to the interparticle 
distance. The confining part of the potential has 
scalar structure \cite{a10,b34} and thereby the interquark potential 
consists of a Coulomb part which is of vector type and a linear 
part which is of scalar type
\begin{equation}
V_{{\rm int}}(r)=V_{{\rm Vector}}(r)+V_{{\rm
Scalar}}(r)= -\frac{4}{3}\frac{\alpha_s(Q^2)}{r}
{[(\gamma^{(1)}_{0}\gamma^{(1)}_{\mu})\otimes
(\gamma^{(2)}_{0}\gamma^{(2)}_{\mu})]}+\kappa r
(\gamma^{(1)}_{0}\otimes\gamma^{(2)}_{0})
\label{funnelaki}
\end{equation}
where $\alpha_s$ is the coupling constant of the strong interaction, 
$Q^2$ is the relevant momentum transfer and $\kappa$ is the string 
constant \cite{a25}.

Certainly, the non-relativistic part of the potential is not 
sufficient if we wish to test the Breit equation. 
The contribution of relativistic corrections are 
required and we must take them into account. At this point, we recall 
that the original Breit equation contained terms called Breit 
interaction that account for relativistic corrections. The inclusion of 
the Breit interaction in the original equation does not lead to 
correct results (\cite{a10}-\cite{a15}) for certain quantum-mechanical systems. Unless 
this term is treated as a small perurbation, the results differ 
from the well-established data.       

By studying the scattering amplitude for a particle-antiparticle 
collision \cite{a10,a30}, we can derive an effective potential in 
the Pauli approximation of the Breit equation. That potential will account 
for the interaction and it will include both 
relativistic and non-relativistic contributions. We need bear in mind that 
the Breit interaction is the second-order term in the $(v/c)$ expansion of 
a Coulomb interaction. In the Pauli approximation, 
the vector part of the quark-antiquark potential contributes 
the following term to the 
interaction 
\begin{eqnarray}
\begin{array}{lcl}
U_{{\rm Vector}}\left(\mbox{\boldmath$\vec{p}$},
\mbox{\boldmath$\vec{r}$}\right)&=&\left\{-\frac{4}{3}
\frac{\alpha_s}{r}\right\}+V_1+V_2+V_3\\
V_1&=&\left\{-\frac{\mbox{\boldmath$\vec{p}$}^4}{8m_1^3}
-\frac{\mbox{\boldmath$\vec{p}$}^4}{8m_2^3}+\frac{2}{3}\pi \alpha_s
\left(\frac{1}{m_1^2}+\frac{1}{m_2^2}\right)\delta
(\mbox{\boldmath$\vec{r}$})\right\}-\frac{2\alpha_s}{3m_1m_2r}
\left[\mbox{\boldmath$\vec{p}$}^2+\frac{\left(\mbox{\boldmath$
\vec{r}$}\cdot \mbox{\boldmath$\vec{p}$}\right)^2}{r^2}\right]\\
V_2&=&\left\{\frac{2\alpha_s}{3r^3}\left[\frac{1}{m_1^2}
\mbox{\boldmath$\vec{L}$}\cdot 
\mbox{\boldmath$\vec{S}$}_1+\frac{1}{m_2^2}
\mbox{\boldmath$\vec{L}$}\cdot 
\mbox{\boldmath$\vec{S}$}_2
\right]\right\}+\frac{4\alpha_s}{3m_1m_2r^3}\mbox
{\boldmath$\vec{L}$}\cdot
\mbox{\boldmath$\vec{S}$}\\
V_3&=&\frac{2\alpha_s}
{m_1m_2r^3}\left[-\frac{1}{3}\mbox{\boldmath$\vec{S}$}^2+
\frac{\left(\mbox{\boldmath$\vec{r}$}\cdot
\mbox{\boldmath$\vec{S}$}\right)
^2}{r^2}\right]+\frac{16\pi \alpha_s }{9
m_1m_2}\left(\mbox{\boldmath$\vec{S}$}^2-\frac{3}{2}\right)\delta  
(\mbox{\boldmath$\vec{r}$})
\label{posaqcd}
\end{array}
\end{eqnarray}
where the $\left\{...\right\}$ terms correspond 
to the static part of the interaction. 
The rest of the terms are 
responsible for retardation effects and they can be associated to the 
Breit interaction. We emphasise that this derivation is in analogy with the 
procedure followed towards obtaining the 
retarded part of the potential in a particle-antiparticle electromagnetic 
interaction (scattering). We recall, that in that case, the contribution of the 
$D_{ik}$ ($i,k=1,2,3$) part of the photon propagator in the 
Coulomb gauge was taken into account \cite{a30}.
 
On the other hand, the scalar confining potential
contributes only to the static part of the interaction. Unlike the vector 
part of the interaction, the scalar term of the potential cannot 
lead to the emergence of retarded effects since there is 
not an equivalent of the photon propagator that would result 
in such a contribution. The scattering
amplitude leads to the following interaction operator \cite{a10,b34}
\begin{eqnarray}
\begin{array}{lcl}
U_{{\rm Scalar}}\left(\mbox{\boldmath$\vec{r}$}\right)&=&
\left\{\kappa r\right\}+V_4\\
V_4&=&\left\{-\frac{1}{2m_1^2m_2^2}\frac{\kappa}{r}
\mbox{\boldmath$\vec{L}$}\cdot \left(m_1^2\mbox{\boldmath$\vec{S}$}_2+
m_2^2\mbox{\boldmath$\vec{S}$}_1\right)\right\}
\label{scalrtfge}
\end{array}
\end{eqnarray}
where $\left\{...\right\}$ merely imply that the interaction is 
characterised by static behaviour. 
For particles with the same mass, $V_2,V_4$ correspond to the spin-orbit
interaction. Although both potentials, $\left(V_{{\rm Vector}}\right)$ and
$\left(V_{{\rm Scalar}}\right)$ are attractive, they give the opposite sign of
spin-orbit interaction, a feature which is going to result in a
quite interesting effect regarding the ordering of the states 
(Section \ref{toblabla}). In addition, we should 
stress the absence of a spin-spin interaction in the confining potential
meaning that tensor forces are contained entirely in the
Coulomb-like vector potential. The absence of spin-spin and tensor terms
in the scalar interaction as well as the different form of spin-orbit
terms between the linear and Coulomb-like potentials constitute the most
striking difference between the two types of interaction.

So far, the form of the potential describing strong 
interactions has been discussed to some extent but nothing has been
mentioned about the efficiency of such a potential. The charmonium and
bottomium spectra are well-known \cite{a35}, therefore the accuracy of the potential
besides the adequacy of the Breit equation in describing strong
interactions can be tested. Next, we will follow a procedure 
towards calculating the energy levels of various bound states of 
quark-antiquark systems, however, we have to obtain first the 
equations. In the following section, we will derive the radial 
equations for those systems as they are produced by means of the 
Eqs.[(\ref{b:a}), (\ref{b:c}), (\ref{b:d})].    

\section{Application of the Breit equation to quarkonium}
\label{sec:four}
The quark-antiquark bound states states can be 
classified in three categories according to their 
spectroscopic signature \cite{a10}: \\
\underline{(i). States with $j=l,S=0$.}
After making the substitutions $V(r)=-\frac{4}{3}\frac{\alpha_s}{r}$, 
$S(r)=-\kappa r$ and eliminating all
components but $\phi^0$, Eqs.(\ref{b:a}) lead to 
\begin{eqnarray}
\frac{d^2\phi^0(r)}{dr^2}&+&\frac{d\phi^0(r)}{dr}\left(\frac{2}{r}-\frac{\kappa
-\frac{4}{3}\frac{\alpha_s}{r^2}}{E+\kappa r
+\frac{4}{3}\frac{\alpha_s}{r}}\right)+\left\{\frac{1}{4}
\left[\left(E+\frac{4}{3}\frac{\alpha_s}{r}\right)^2-
\kappa ^2r^2\right] \right.- \nonumber \\
&-&\left. \frac{j(j+1)}{r^2}-m^2\frac{E+\kappa r+
\frac{4}{3}\frac{\alpha_s}{r}}{E-\kappa r+
\frac{4}{3}\frac{\alpha_s}{r}}\right\}\phi^0(r)=0
\label{thaxrh1}
\end{eqnarray}
\underline{(ii). States with $j=l,S=1$.}\\
Similarly, Eqs.(\ref{b:c}) give 
\begin{eqnarray}
\frac{d^2\chi_m^0(r)}{dr^2}&+&\frac{d\chi_m^0(r)}{dr}\left(\frac{2}{r}-
\frac{\kappa- \frac{4}{3}\frac{\alpha_s}{r^2}}{E+\kappa r
+\frac{4}{3}\frac{\alpha_s}{r}}\right)+\left\{\frac{1}{4}
\left[\left(E+\frac{4}{3}\frac{\alpha_s}{r}\right)^2-
\kappa ^2r^2\right]-\frac{j(j+1)}{r^2}-\right. \nonumber \\
&-&\left. m^2\frac{E+\kappa r+            
\frac{4}{3}\frac{\alpha_s}{r}}{E-\kappa r+
\frac{4}{3}\frac{\alpha_s}{r}}-\frac{1}{r}\frac{\kappa -
\frac{4}{3}\frac{\alpha_s}{r^2}}{E+\kappa r+
\frac{4}{3}\frac{\alpha_s}{r}}\right\}\chi_m^0(r)=0
\label{thaxrh2}
\end{eqnarray} 
\underline{(iii). States with $j=l\pm 1,S=1$.}\\
Finally, Eqs.(\ref{b:d}) can be reduced to the following two coupled
differential equations 
\begin{eqnarray}
&&\frac{d^2\chi^0_e(r)}{dr^2}+\frac{d\chi^0_e(r)}{dr}\left[\frac{2}{r}
+\frac{-\frac{4\alpha_s}{3r^2}-\kappa}{E+\frac{4\alpha_s}{3r}-
\kappa r}+\frac{-2\kappa \left(E+\frac{8\alpha_s}{3r}\right)}
{\left(E+\frac{4\alpha_s}{3r}\right)^2-(\kappa r)^2}\right]+ \nonumber \\
&\mbox{      }&+\chi^0_e(r) 
\left\{\frac{\left(E+\frac{4\alpha_s}{3r}+
\kappa r\right)\left[\frac{1}{4}\left(E+\frac{4\alpha_s}{3r}-
\kappa r\right)^2-m^2
\right]}{E+\frac{4\alpha_s}{3r}-
\kappa r}-\frac{j(j+1)}{r^2}+
\right .  \nonumber \\
&\mbox {      }&\left. +\frac{-4\kappa
\left(E+\frac{8\alpha_s}{3r}\right)}{\left(E+\frac{4\alpha_s}{3r}\right)^2-
(\kappa
r)^2}\frac{1}{r}-\frac{2}{r^2}+\frac{\kappa^2 +\frac{16\alpha_s^2}{9r^4}+
\frac{8\alpha_s\kappa }{3r^2}+\frac{8\alpha_sE}{3r^3}}{E+\frac{4\alpha_s}
{3r}+  
\kappa r} \frac{1}{E+\frac{4\alpha_s}{3r}-
\kappa r}\right\}= \nonumber \\
&\mbox{      }&=\chi_l^0(r)\left[\frac{1}{r^2}-\frac{-2\kappa
\left(E+\frac{8\alpha_s}{3r}\right)}{\left(E+\frac{4\alpha_s}{3r}+
\kappa r\right)^2}\frac{1}{r}+\frac{\left(E+\frac{4\alpha_s}{3r}-
\kappa r\right)}{\left(E+\frac{4\alpha_s}{3r}+
\kappa r\right)}\frac{1}{r^2}
\right]
\label{thaxrh3}
\end{eqnarray}
\begin{eqnarray}
&&\frac{d^2\chi^0_l(r)}{dr^2}+\frac{d\chi^0_l(r)}{dr}\left[\frac{2}{r} 
+\frac{-\frac{4\alpha_s}{3r^2}-\kappa}{E+\frac{4\alpha_s}{3r}+
\kappa
r}\right]+\chi^0_l(r)\left\{\frac{E+\frac{4\alpha_s}{3r}+
\kappa r}{E+\frac{4\alpha_s}{3r}-
\kappa r}\times 
\right .  \nonumber \\
&\mbox{      }&\left.\times \left[\frac{1}{4}\left(E+\frac{4\alpha_s}{3r}
-\kappa r\right)^2-m^2
\right]-\frac{j(j+1)}{r^2}-\frac{1}{r}\frac{-\frac{4\alpha_s}{3r^2}+
\kappa}{E+\frac{4\alpha_s}{3r}+
\kappa r} \right\}= \nonumber \\
&\mbox{      }&=\chi^0_e(r)\left[\frac{2j(j+1)}
{r^2}+\frac{j(j+1)}{r}\left(\frac{-
\frac{4\alpha_s}{3r^2}-\kappa}{E+\frac{4\alpha_s}{3r}-
\kappa r}+\frac{-
\frac{4\alpha_s}{3r^2}+\kappa}{E+\frac{4\alpha_s}{3r}+
\kappa r}\right)
\right]
\label{thaxrh4}
\end{eqnarray}
The behaviour of the component wavefunctions in the 
above equations near the origin can be deduced easily if we 
notice that at very small interparticle distances, the dominant part of the 
potential has a vector, Coulomb-like form. 
Thus, the wavefunctions for small $r$ look like 
\begin{eqnarray}
\begin{array}{lcl}
{\rm state}\mbox{ } ^1l_l\hspace{.6cm}, 
&r\phi^0\sim r^{\gamma +1} &, \hspace{.4cm} \gamma=-1+
\sqrt{1+j(j+1)-\frac{4}{9}\alpha^2_s}\\
{\rm state}\mbox{ } ^3l_l\hspace{0.6cm}, 
&r\chi_m^0\sim r^{\gamma +1} &, \hspace{.4cm} \gamma=-1+
\sqrt{j(j+1)-\frac{4}{9}\alpha^2_s}\\
{\rm state}\mbox{ } ^3l_{l\pm 1}\hspace{.2cm},&\left\{
\begin{array}{l}
r\chi^0_e \sim r^{\gamma+1}\\
r\chi^0_l \sim r^{\gamma+1}
\end{array}\right\}& ,\hspace{.4cm} \gamma=\left\{
\begin{array}{l}
\sqrt{j(j+1)+1-\frac{4}{9}\alpha^2_s }\\
-1+\sqrt{j(j+1)-\frac{4}{9}\alpha^2_s }
\end{array}\right. 
\label{whichgamma}
\end{array}
\end{eqnarray} 
On the other hand, due to the inclusion of the confining potential, the
wavefunctions have a different behaviour at large distances. They look
like $\sim e^{-\frac{1}{4}\kappa r^2}$. The behaviour of the wavefunctions near the origin and at large distances will be used in the next sections in order to solve the differential equations describing the dynamics of the quantum mechanical systems. 

\section{\bf{Solution of the equations}}
\label{sec:five}
As pointed out in the previous sections, 
the energy levels of the various states are derived from the solution of
the above radial equations, however, those values do not represent the
complete energy since the potential used is merely the static potential. 
The contribution of the retarded part of the
interaction should be included to obtain the relativistic corrections
to the energy. The static part consists of the $\{...\}$ terms 
in the expressions [(\ref{posaqcd}), (\ref{scalrtfge})] while the rest are
the retarded terms. At this point, we should recall that we do not expect 
satisfactory results unless the Breit (retarded) terms are treated by 
first order perturbation theory for particular states. The application to 
QED problems (\cite{a10}-\cite{a15}) demonstrated this assertion is correct. In that case, 
the total energy of the system is given by the 
expression 
\begin{eqnarray}
E_{{\rm total}}=E_{{\rm static}}+\left<{\rm state}\left|V_{{\rm retarded}}
\right|{\rm state}\right>
\end{eqnarray}
where $\left.\left|{\rm state}\right.\right>$ stands for the stationary
unperturbed states for the Coulomb potential. The same procedure can be 
followed when the potential has the form (\ref{funnelaki}), however,
unlike QED problems, the eigenstates are not known and they
should be calculated before we proceed. Unfortunately, an exact solution
for such a potential cannot be obtained and an approximate method should
be tried. One way is by treating the Coulomb term as a small
perturbation \cite{a40}. Another method, which will be employed in 
this paper, is
to use the three-dimensional isotropic harmonic oscillator (TDIHO) 
eigenstates \cite{a10,a45}. The
potential of the harmonic oscillator provides confinement of quarks and
the important feature of this potential is that the wavefunctions and
the matrix elements can be calculated easily and they can possess an 
explicit form.
The parameters of the wavefunctions will be adjusted so as to fit the
numerically calculated eigenfunctions for the potential
(\ref{funnelaki}). 

The interaction potential of TDIHO equals 
\begin{eqnarray}
V_{{\rm TDIHO}}=\frac{1}{2}\mu \omega^2r^2
\end{eqnarray}
where $\mu$ and $\omega$ are fitting parameters. The energy levels of the
system are given by the expression
\begin{eqnarray}
E(n_r,l)=\left(2n_r+l+\frac{3}{2}\right)\omega, \hspace{0.5cm} n_r,l=0,1,2,...
\end{eqnarray}
and the normalised radial parts of the wavefunctions which are going to
be used
are the following \cite{a45}
\begin{eqnarray}
\begin{array}{lcl}
R_{1S}=2\underbrace{\left(\frac{\lambda_{1S}^3}{\pi}\right)^{1/4}\exp
\left(-\frac{\lambda_{1S}}{2}r^2\right)}_{B_{1S}(r)} &, & R_{1P}=\sqrt{\frac{8
\lambda_{1P}}{3}}rB_{1P}(r) \nonumber \\
R_{2S}=\sqrt{6}
\left(1-\frac{2}{3}\lambda_{2S} r^2\right)B_{2S}(r) &,&
R_{2P}=\sqrt{\frac{4\lambda_{2P}}{15}}r\left(5-2
\lambda_{2P} r^2\right) B_{2P}(r) \nonumber \\
R_{3S}= \frac{1}{\sqrt{30}}\left(15-20\lambda_{3S} r^2+4
\lambda^2_{3S}r^4\right)B_{3S}(r)&, &
R_{4S}=\frac{1}{\sqrt{1260}}\left(105-210\lambda_{4S} r^2+
\right.\nonumber \\
&&\left. \hspace{1cm} +84
\lambda^2_{4S}r^4-8\lambda_{4S}^3r^6\right) B_{4S}(r) \nonumber \\
R_{5S}=\frac{1}{\sqrt{90720}} \left(945-2520\lambda_{5S} r^2+1512 
\lambda_{5S}^2r^4- \right. & &R_{6S}=
\frac{1}{\sqrt{9979200}}\left(10395-34650 \lambda_{6S} r^2+
\right.\nonumber \\
\left. \hspace{1cm}  -288 
\lambda_{5S}^3 r^6+16 \lambda_{5S}^4 r^8\right)B_{5S}(r)&, & 
\left. \hspace{1cm} +27720 
\lambda_{6S}^2 r^4-
7920 \lambda_{6S}^3 r^6+ \right. \nonumber \\
&&\left. \hspace{1cm}+ 880 \lambda_{6S}^4 r^8-32 
\lambda_{6S}^5 r^{10}\right)B_{6S}(r) \nonumber
\end{array}
\end{eqnarray}
where $B_{\rm index}\equiv B\left(\lambda_{1S}\rightarrow 
\lambda_{{\rm index}}
\right)$. $\lambda_{1S}\equiv \lambda =\mu 
\omega$ while  the rest of the $\lambda$s
are functions of $\lambda_{1S}$ and their values will be determined from 
experimental data. This is done because the harmonic oscillator is
an approximation of the ``funnel'' potential \cite{b1,b20} 
therefore the values of the
various parameters need to be adjusted to be in agreement with the
experimental values of the corresponding quantities. It is necessary to
emphasise that the $\lambda$s have a mass 
dependence implying that the relation among them which will emerge
is not the same for all quarkonium systems. The parameters 
$\mu, \omega$, in turn,  can be estimated by comparing the values of the
quantities in which they enter with those derived from experimental
data. The leptonic widths, the mass differences and the various energy
levels of some certain states are quantities that can help us
to estimate not only those parameters but also other parameters such as the
coupling constant $\alpha_s$, the string constant $\kappa$ and the quark
masses. To achieve this goal, we will take advantage of the knowledge
of the energy levels of the charmonium and bottomium systems \cite{a35}.

\subsection{\bf{The Bottomium system}}
\label{toblabla}
We aim to estimate the energies of the twelve states of the bottomium
system which are spin triplet states,
i.e. $S=1$
({\it ortho-bottomium}). 
The $1^{--}, 0^{++}, 2^{++}$ states are vector mesons and the
static part of the interaction between the constituent particles leads
to the description of their dynamics by Eqs.[(\ref{thaxrh3}), 
(\ref{thaxrh4})]. On the
other hand, the rest of the
states, denoted with $1^{++}$, are pseudoscalar mesons and they are 
described by Eq.(\ref{thaxrh2}). Finally, 
six more pseudoscalar meson states (Table \ref{taeftmore})
(with $J^{PC}=0^{-+}, 1^{+-}$) will be considered
and Eq.(\ref{thaxrh1}) describes those states. Unlike QED, 
the explicit form of the potential of QCD and the coupling strength are
not known therefore a method towards estimating them should be
formulated. The procedure becomes more difficult if we remember that the
solution of the above equations is not enough to get the energy spectrum
since the values calculated account only for a part of the whole energy.
The rest of the contribution results from relativistic effects due to
the non-instantaneous interaction which 
should not be ignored. The combination leads to the problem resisting an
analytic treatment and only a numerical method seems capable of solving it.

Since the coupling constant $\alpha_s$ of the system as well as the mass of the
bottom quark $m_b$ are not known, the use of some of the states to determine these
values is inevitable. In addition, there are some
more parameters which need to be estimated: the string constant $\kappa$
and the $\lambda$s. Before we embark on calculating the various
quantities, we will first attempt to express the energies of the 
states in terms of the unknown parameters.

The energy of the system consists of two parts, one coming from the
non-perturbative solution of the Eqs.[(\ref{thaxrh1})-(\ref{thaxrh4})] 
and another derived from the perturbative
treatment of the non-retarded parts of 
$V_1,V_2,V_3$ in (\ref{posaqcd}) by repacing $m_1, m_2$
with $m_b\equiv m$. The former will be
called $E_{\rm{static}}$ while the latter which is non-instaneous, in
nature, will be called $E_{\rm{non-static}}$. The combined
result in the total energy of the system $E_{\rm{system}}$ is 
\begin{eqnarray}
E_{\rm{system}}&=&E_{\rm{static}}+\underbrace{\left<V_1+V_2+V_3\right>}_
{E_{\rm{non-static}}}=E_{\rm{static}}+\left<-\frac{2\alpha_s}
{3m^2}\left[-\frac{2}{r}\frac{d^2}{dr^2}-\frac{2}{r^2}\frac{d}{dr}+
\frac{1}{r^3}\mbox{\boldmath$\vec{L}$}^2\right] \right.+ \nonumber \\
&\mbox{      }&\left.+ 
\frac{4\alpha_s}{3m^2r^3}\mbox{\boldmath$\vec{L}$}\cdot 
\mbox{\boldmath$\vec{S}$}+\frac{2\alpha_s}{m^2r^3}
\underbrace{\left[-\frac{1}{3}\mbox{\boldmath$\vec{S}$}^2+
\frac{\left(\mbox{\boldmath$\vec{r}$}\cdot
\mbox{\boldmath$\vec{S}$}\right)
^2}{r^2}\right]}_{S_{12}}+
\frac{16\pi\alpha_s}{9m^2}\left(\mbox{\boldmath$\vec{S}$}^2-\frac{3}{2}
\right)\delta\left(\mbox{\boldmath$\vec{r}$}\right)
\right>
\label{thasjusd}
\end{eqnarray}
The spin-orbit perturbative correction and $S_{12}$ vanish for
$S$-states while the $\delta$ functions give a non-zero contribution. By
contrast, the $P$-states are characterised by the opposite behaviour
$\left( \mbox{\boldmath$\vec{L}$}\cdot \mbox{\boldmath$\vec{S}$},
S_{12}\neq 0,
\left<\delta\left(\mbox{\boldmath$\vec{r}$}\right)\right>=0
\right)$. In Table \ref{taeft}, the energies of various states are summarised
and we notice that the knowledge of the $\lambda$s is necessary to obtain 
the values of the corrections due to the Breit terms.
The decay widths of the states will be used, next, to evaluate the
relation of the $\lambda$s and other parameters to some measured
quantities. 
The ``Corrected Van Royen-Weisskopf'' formula \cite{b5,b14} 
\begin{eqnarray}
\Gamma\left(V\rightarrow e^+e^-\right)=\frac{16\pi
\alpha^2 Q_b^2}{m^2_V}\left|\psi(0)\right|^2\left(1-
\frac{16}{3\pi}\alpha_s\right),\hspace{0.5cm} \alpha=\frac{1}{137},
\hspace{0.5cm} Q_b=\frac{1}{3}
\label{decaywid}
\end{eqnarray}
relates the leptonic width, i.e. $e^+e^-$ decay of the neutral $V$ vector
mesons, and the wavefunction of bottomium at the origin. $m_V$ stands
for the mass of the meson. 
Although the
coupling constant has not been estimated yet, the fact that we are dealing
with strong interactions suggests that the radiative corrections are so
large that only ratios such as 
\begin{eqnarray}
r(V'/V)\equiv \frac{\Gamma\left(V'\rightarrow e^+e^-\mbox{ }\rm or
\mbox{ }\mu^+\mu^-
\right)}{\Gamma 
\left(V\rightarrow e^+e^-\mbox{ }\rm or \mbox{ }\mu^+\mu^-
\right)}=\frac{m^2_V}{\left(m_V'\right)^2}\frac{\left|\psi'(0)\right|^2}
{\left|\psi(0)\right|^2}
\label{gammaratio}
\end{eqnarray}
can be calculated reliably because the corrections are suppressed and
do not appear in the above expressions. In (\ref{gammaratio}), $V'$ is
another vector state of the quarkonium system having mass $m'_V$. All
$1^{--}$ vector states can decay into an electron-positron (or
muon-antimuon) pair and the
five ratios of these decays with respect the ``reference decay'' of the
$\Upsilon (1S)$ state are given from experiment \cite{a35}
\begin{eqnarray}
r(2S/1S)_{\mu}&\equiv&\frac{E_{\Upsilon(1S)}^2}{E_{\Upsilon(2S)}^2}
\frac{\left|\psi_{\Upsilon(2S)}(0)\right|^2}
{\left|\psi_{\Upsilon(1S)}(0)\right|^2}=0.4\pm 0.1  \label{rat2}   \\
r(3S/1S)_{\mu}&\equiv&\frac{E_{\Upsilon(1S)}^2}{E_{\Upsilon(3S)}^2} 
\frac{\left|\psi_{\Upsilon(3S)}(0)\right|^2}
{\left|\psi_{\Upsilon(1S)}(0)\right|^2}=0.37\pm 0.06  \label{rat3} \\
r(4S/1S)&\equiv&\frac{E_{\Upsilon(1S)}^2}{E_{\Upsilon(4S)}^2}
\frac{\left|\psi_{\Upsilon(4S)}(0)\right|^2}
{\left|\psi_{\Upsilon(1S)}(0)\right|^2}=0.18 \pm 0.04  \label{rat4} \\
r(5S/1S)&\equiv&\frac{E_{\Upsilon(1S)}^2}{E_{\Upsilon(5S)}^2}
\frac{\left|\psi_{\Upsilon(5S)}(0)\right|^2}
{\left|\psi_{\Upsilon(1S)}(0)\right|^2}=0.23\pm 0.05  \label{rat5} \\
r(6S/1S)&\equiv&\frac{E_{\Upsilon(1S)}^2}{E_{\Upsilon(6S)}^2}
\frac{\left|\psi_{\Upsilon(6S)}(0)\right|^2}
{\left|\psi_{\Upsilon(1S)}(0)\right|^2}=0.10\pm 0.02
\label{rat6}
\end{eqnarray}
(the index $\mu$ in (\ref{rat2}) and (\ref{rat3}) indicates that the vector
mesons involved decay into muon-antimuon pairs). Each one of $r(nS/1S)$ 
($n=2,...6)$ is a function of the energy of the system and two
$\lambda$s, i.e., $\lambda_{1S}$ and $\lambda_{nS}$. 

The relations [(\ref{rat2})-(\ref{rat6})]
can be used to
establish the relation among the $\lambda$s, energies and $m$.
(\ref{thasjusd}) is another useful expression which will be employed in
order to calculate the energies of the various states as well as the
values of $\alpha_s, \kappa, m$. The part of the energy due to the
instantaneous interaction should be combined with the non-retarded 
corrections so that the energy obtained is in agreement with experiment.
This is called ``fitting procedure'' and it will help us to determine the
four unknown parameters $\alpha_s, \kappa, m, \lambda_{1S}$. The rest of
the $\lambda$s, associated with the $S$-states, 
are related to them through the decay widths $\Gamma\left(V\rightarrow
e^+e^- \right)$ 
whose values
are well established. There are two more $\lambda$s, namely, 
$\lambda_{1P}$ and $\lambda_{2P}$ which cannot be calculated accurately
by means of the transitions decays, as pointed out in the previous
paragraph. 
The total number of the unknown parameters, i.e. six,
requires six energy values to be set as input values. 
We choose the
energies of the $\Upsilon(1S), \Upsilon(2S), \Upsilon(3S), 
\Upsilon(4S),  
\chi_{b1}(1P), \chi_{b1}(2P)$ states as the input values. Although 
$P$-states with the same principal quantum number 
can be used, we avoid it because they correspond to energies very close
to one another which may lead to wrong results. The 
harmonic oscillator wavefunctions constitute an approximation to the
real wavefunctions of the funnel potential and in the stage of
optimising them, the method may be too sensitive to small 
energy differences. Thus, we choose the following values established by
experiment \cite{a35}   
\begin{eqnarray}
\begin{array}{lcl}
E\left[\Upsilon(1S)\right]&=& 9460.37\pm 0.21 \mbox{ }\rm MeV\\
E\left[\Upsilon(2S)\right]&=& 10023.30\pm 0.31 \mbox{ }\rm MeV\\
E\left[\Upsilon(3S)\right]&=& 10355.3\pm 0.5 \mbox{ }\rm MeV\\
E\left[\Upsilon(4S)\right]&=& 10580.0\pm 3.5 \mbox{ }\rm MeV\\
E\left[ \chi_{b1}(1P)\right]&=&9891.9\pm 0.7 \mbox{ }\rm MeV \\
E\left[ \chi_{b1}(2P)\right]&=&10255.2\pm 0.5 \mbox{ }\rm MeV \\
\end{array}
\label{inputval}
\end{eqnarray}  
The energies of the $\Upsilon$ states will help us to determine the 
relations between the $\lambda_{nS}$ $(n=2,3,4)$ and $\lambda_{1S}$
through the ratios [(\ref{rat2})-(\ref{rat4})].
 
The mass of the bottom quark is regarded as a 
parameter which will be evaluated, however, its value will be 
contstrained 
in the range $4.1\mbox{ }\rm {GeV \leq}$ $m_b$ $\rm {\equiv}$ $m$ 
$\leq 5\mbox{ }\rm 
GeV$
(running mass) \cite{a35}. 
The energies due to the static interaction can be expressed easily in
terms of the total energy and the parameters. From (\ref{thasjusd}),
$E_{\rm static}=E_{\rm total}-E_{\rm perturbation}$ which means that the
numerical solution of Eqs.[(\ref{thaxrh2})-(\ref{thaxrh4})] can lead to
the determination of the parameters, provided that the relations
connecting them dictated by the corresponding quantities are satisfied.
Thus, the main problem reduces to the solution of the differential
equations. Certainly, it is important to observe that Eqs. 
[(\ref{thaxrh1})-(\ref{thaxrh4})] exhibit a singularity at 
\begin{equation}
r_0=\frac{1}{2}\frac{E}{\kappa}+\sqrt{\frac{1}{4}\left(\frac{E}
{\kappa}\right)^2+\frac{4}{3}\frac{\alpha_s}{\kappa}}
\label{jdfsing}
\end{equation}
This is a feature which is not present in the initial Breit equation but it emerges after the reduction of it to radial equations. Although the equations are singular at $r=0$ as well, due to the centrifugal 
term $\frac{j(j+1)}{r^2}$ ({\it j} is the total angular momentum of the
system), the new singularity implies the appearance of a turning point that is energy dependent \cite{a10,a13,a48}. The question rising is whether the emergence of this singularity causes any problem to the solution of the differential equations. Due to the fact that we consider only 
a short-range potential and the wavefunction tends to vanish at 
distances ($r\geq 1\mbox{ }\rm  fm$), we will not experience any difficulty in solving the differential equations. It will be shown that
$r_0$ is bigger than $10\mbox{ }\rm fm$ (for the lowest energy) which
implies that the singularity 
does not cause any problem in the range where QCD applies, on the grounds
that the equations have meaning in this particular range.    

The $\chi$ and $\Upsilon$ 
states of the ``fitting procedure'' are described by
Eq.(\ref{thaxrh2}) and Eqs.((\ref{thaxrh3}), (\ref{thaxrh4})), 
respectively, and these
equations can be reduced to a set of twenty, first order differential
equations. The problem is a boundary-value problem and two boundary
points, $R_{{\rm initial}}, R_{{\rm final}}$ 
need to be specified between 
which the integration will take place. We choose $R_{{\rm final}}
\simeq 3 \mbox{ }\rm fm$
where the wavefunction and its derivative are taken
to vanish due to the
confinement. On 
the other hand, the choice of $R_{{\rm initial}}$
requires more care. At small distances, the Coulomb potential dominates
the interaction and since the potential is singular at $r=0$, we are not
allowed to start the integration at that point, therefore we choose a
point which is much smaller than $r_c\equiv \frac{\alpha_s}{\Lambda}$
($\Lambda$: QCD scale parameter). The QCD scale parameter $\Lambda$
is not an independent parameter, it is related, instead, to 
$\kappa$ \cite{b25,b34} through the expression 
\begin{eqnarray}
\kappa=\frac{8\pi \Lambda^2}{27}
\label{kapalam}
\end{eqnarray}
We set $R_{{\rm initial}}=10^{-10}r_c$. The
wavefunction behaves as $r^{\gamma}$ at $r=R_{{\rm initial}}$ while it
vanishes at $r=R_{{\rm final}}$. According
to (\ref{whichgamma}), 
\begin{eqnarray}
\gamma=\left\{
\begin{array}{lcl}
-1+
\sqrt{j(j+1)-\frac{4}{9}\alpha^2_s}&,&\hspace{1cm}{\rm for}\mbox{ }\chi_{b1}
 \mbox{ }\rm states\\
\sqrt{j(j+1)+1-\frac{4}{9}\alpha^2_s}&,&\hspace{1cm}{\rm for}\mbox{ }
\Upsilon, \chi_{b0}, \chi_{b2} \mbox{ }\rm states
\end{array}
\right.
\end{eqnarray}
The reason why the bigger value of $\gamma$ was preferred 
in the $^3l_{l\pm 1}$ states is because the wavefunction goes to zero
faster.

The numerical integration of the differential equations is achieved by
means of a Runge-Kutta-based routine and it leads to the
following values for the parameters \cite{a10}
\begin{eqnarray}
\begin{array}{lcl}
\alpha_s  &=&  0.36  \pm 0.04 \\
\kappa    &=&  0.793 \pm 0.008\mbox{ } \frac{\rm GeV}{\rm fm}\\
m_b\equiv m &=&  4.987 \pm 0.009 \mbox{ } \rm GeV\\
\lambda_{1S} &\simeq &  0.829 \mbox{ } ({\rm GeV})^2\\
\lambda_{1P} &\simeq&  0.107 \mbox{ } ({\rm GeV})^2  \\
\lambda_{2P} &\simeq&  0.120 \mbox{ } ({\rm GeV})^2 
\end{array}
\end{eqnarray}
which, in turn, give 
\begin{eqnarray}
\begin{array}{lcl}
\lambda_{2S}  &\stackrel{(\ref{rat2})}{\simeq}&   
0.147 \mbox{ } ({\rm GeV})^2   \\
\lambda_{3S}   &\stackrel{(\ref{rat3})}{\simeq}&    
0.126 \mbox{ }  ({\rm GeV})^2  \\
\lambda_{4S}   &\stackrel{(\ref{rat4})}{\simeq}&   
0.072 \mbox{ } ({\rm GeV})^2 \\ 
\Lambda &\stackrel{(\ref{kapalam})}{=}& 409\pm 40 \mbox{ } \rm MeV
\end{array}
\end{eqnarray}
 
Now that the parameters have been estimated, they can be used to
calculate the rest of the energy levels of the system. The states that 
are going to be considered can be classified in three groups according
to the set of equations that describe their instantaneous behaviour or, 
equivalently, the spectroscopic signature: (i). $\eta(1S), \eta(2S),
\eta(3S),\eta(4S)$, $h_b(1P), h_b(2P)$, with $^{2S+1}L_J\equiv 
^1l_l$, 
(ii).$\chi_{b1}(1P), \chi_{b1}(2P)$, with $^{2S+1}L_J\equiv^3l_l$, and, 
finally,  
(iii). $\Upsilon(1S), \Upsilon(2S),\Upsilon(3S),
\Upsilon(4S),
\Upsilon(5S),\Upsilon(6S), \chi_{b0}(1P),\chi_{b0}(1P),
\chi_{b0}(2P),\chi_{b2}(2P)$, with $^{2S+1}L_J\equiv^3l_{l\pm 1}$. By
applying Eqs.[(\ref{thaxrh1})-(\ref{thaxrh4})] 
to the cases (i), (ii) and (iii), and by  
imposing the same boundary conditions, as previously, the energies of the
states can be obtained. The $\lambda$s of the wavefunctions entering the
equations are dependent on the energies of the corresponding states,
therefore the expressions [(\ref{rat5}), (\ref{rat6})] should be employed. 
In addition, Table \ref{taeft} and Table \ref{taeftmore} provide the
contribution of the Breit terms to the energy of the system. If all
these elements are taken into account, the energy values of the 
above states are calculated and they are summarised in Table
\ref{sumall} and Table \ref{sumallaa}. The experimentally established energies are also stated.   

The two remaining $\lambda$s, namely, $\lambda_{5S}, \lambda_{6S}$ are
calculated and they are equal to 
\begin{eqnarray}
\begin{array}{lcl}
\lambda_{5S} &\simeq& 0.082 \mbox{ } ({\rm GeV})^2 \\
\lambda_{6S} &\simeq & 0.045 \mbox{ } ({\rm GeV})^2 
\end{array}
\end{eqnarray}
Looking at Table \ref{sumall} and Table \ref{sumallaa}, there are
some points which need to be emphasised:\\
(i). we notice that the states predicted from the theory have the right
accession and almost 
the correct energy difference between the $P$ states
emerges. Actually, there is not
a very big discrepancy between the theoretical results and the experimental
data as it can be deduced from Table \ref{tablesplit}. 
For higher states, it is getting bigger and the discrepancy becomes larger and 
this is 
quite reasonable since those states are characterised by large speed, 
therefore the $v^2$ approximation to the Breit terms constitutes a rather
rough approximation,  \\
(ii). the part of the energy due to the instantaneous interaction
accounts for more of the energy, however, the contribution of terms due to 
retardation is very important
to acquire the correct splitting of the $P$ states. At this point, an
interesting feature of the funnel potential should be mentioned. The
spin-orbit interaction coming from the Coulomb-like part of the
potential (\ref{posaqcd}) differ from the spin-orbit interaction 
resulting from the linear
part (\ref{scalrtfge}) in the sign. Thus, if only the Coulomb-like
potential was present, it would order the $P$-levels, in ascending order
$^3P_0$, $^3P_1$, $^3P_0$. On the other hand, the existence solely of
the scalar part would order them oppositely. The fact that the previous
order is obtained suggests that the spin-orbit part due to the vector
potential dominates,     \\
(iii). the ratios $\rho\left(\chi_{bJ}(1P)\right), 
\rho\left(\chi_{bJ}(1P)\right)$  
$\left(\equiv (2^{++}-1^{++})/(1^{++}-0^{++})\right) $can be 
calculated from Table \ref{tablesplit}. They are equal to 
\begin{eqnarray}
\begin{array}{lcl}
\rho\left(\chi_{bJ}(1P)\right)  &=0.69 \pm 0.03& \nonumber \\ 
\rho\left(\chi_{bJ}(2P)\right)  &= 0.56 \pm 0.09 & \nonumber 
\end{array}
\end{eqnarray}
which are in a very good agreement with experiment \cite{a35}
\begin{eqnarray}
\begin{array}{lcl}
\rho\left(\chi_{bJ}(1P)\right)  &=0.67 \pm 0.03& \nonumber \\ 
\rho\left(\chi_{bJ}(2P)\right)  &= 0.58 \pm 0.03 & \nonumber 
\end{array}
\end{eqnarray}
(iv). the states are characterised by small speed
$\left(\left<(v/c)^2\right>\leq 
0.11\right)$ which make the application of the Pauli
approximation and perturbation theory possible.
$\left<\left(v/c\right)^2\right>$ is proportional to the $\lambda$ of
the corresponding state and inversely proportional to the mass of the
bottom quark,  \\
(v). $r_0$ depends on the energy of the state and
$r_0\simeq 12 \mbox{ }\rm fm$. $\left[\right.R_{\rm initial}, 
r_0\left.\right)$ that is
clearly larger than the range within which the interaction takes place.
If we continue the integration of the differential equations for bigger
interparticle distances, we notice that at $r=r_0$
both the wavefunction and its derivative vanish due to confinement. \\
(vi). the splittings of the $1S, 2S, 3S, 4S$ states are given from
Table \ref{tablesplits}.

The successful description of bottomium by means of the Breit equation
can serve as a first indication that the equation can be used in
short-range interactions, however, before we generalise, it would be wise
to study the spectrum of other quarkonia composed of lighter quarks.

\subsection{\bf{The Charmonium system}}
\label{sec:nine}
In this section, a similar procedure is going to be followed in order 
to calculate the charmonium spectrum. The twelve states we will consider
have the quantum numbers $J^{PC}=1^{--}, 0^{++},2^{++}, 0^{-+},1^{+-}$.
As in the previous subsection, the widths of the decays of 
the $S$ states into $e^+e^-$ will be employed and 
the relations between the $\lambda$s of the
radial part of the wavefunctions will be eventually revealed. 
The decays lead to ratios given 
by the following expressions  
\begin{eqnarray}
r(2S/1S)&\equiv&\frac{E_{J/\psi(1S)}^2}{E_{\psi(2S)}^2}  
\frac{\left|\psi_{\psi(2S)}(0)\right|^2}  
{\left|\psi_{J/\psi(1S)}(0)\right|^2}=0.41\pm 0.05  \label{rat2c}   \\
r(3S/1S)&\equiv&\frac{E_{J/\psi(1S)}^2}{E_{\psi(3S)}^2}
\frac{\left|\psi_{\psi(3S)}(0)\right|^2}
{\left|\psi_{J/\psi(1S)}(0)\right|^2}=0.14\pm 0.03  \label{rat3c} \\
r(4S/1S)&\equiv&\frac{E_{J/\psi(1S)}^2}{E_{\psi(4S)}^2}
\frac{\left|\psi_{\psi(4S)}(0)\right|^2}
{\left|\psi_{J/\psi(1S)}(0)\right|^2}=0.09 \pm 0.02  \label{rat4c}
\end{eqnarray}
The parameters that need to be evaluated are three, the mass of the charm
quark, $m_c\equiv m$, $\lambda_{1S}$ and $\lambda_{1P}$. It is assumed
that the coupling constant $\alpha_s$ and string
constant
$\kappa$ are the same as in the bottomium system. 
This lies in the so-called {\it flavour independence} \cite{a50} 
of the strong
interactions which suggests that the interaction between $c,\bar{c}$ does
not differ from that between $b, \bar{b}$. The same holds true for any
combination of two quarks. Since all quarks exist in the same 
three colour states,
they must have identical strong interactions. 

The ``fitting procedure'' \cite{a10} is going to be followed, in order to
determine the three parameters. We first impose the constraint that the
mass of the charm quarks lies within the range $1\mbox { } \rm GeV \leq
$ $m$ $\leq 1.6 \mbox{ }\rm GeV$ (running mass) \cite{a35}. The energies of
the states $J/\psi(1S), \psi(2S)$, $\chi_{c1}(1P)$ are chosen as input
values and the ratios [(\ref{rat2c})-(\ref{rat4c}]) 
as well as the differential equations
Eqs.[(\ref{thaxrh2})-(\ref{thaxrh4})] are used. The latter are
integrated in the range $\left[10^{-10}r_c,3\mbox{ }\rm fm\right]$
where $r_c=\alpha_s/\Lambda$. 
By solving the differential equations 
and by taking into account the relations between the 
parameters imposed
by [(\ref{rat2c})-(\ref{rat4c})], we obtain the following results \cite{a10}
\begin{eqnarray}
\begin{array}{lcl}
m_c\equiv&=& 1.572 \pm 0.009 \mbox{ }\rm GeV\\
\lambda_{1S}&\simeq &0.395 \mbox{ } (\rm GeV)^2\\
\lambda_{1P}&\simeq &0.074 \mbox{ } (\rm GeV)^2
\end{array}
\end{eqnarray}
which, in turn, give
\begin{eqnarray}
\begin{array}{lcl}
\lambda_{2S}  &\stackrel{(\ref{rat2c})}{\simeq}&
0.083 \mbox{ } ({\rm GeV})^2   \\
\lambda_{3S}   &\stackrel{(\ref{rat3c})}{\simeq}&
0.040 \mbox{ }  ({\rm GeV})^2  \\
\lambda_{4S}   &\stackrel{(\ref{rat4c})}{\simeq}&
0.030 \mbox{ } ({\rm GeV})^2 \\
\end{array}
\end{eqnarray}

Now that the parameters have been estimated, they can be used to   
calculate the rest of the energy levels of the system. The application
of Eqs.[(\ref{thaxrh1})-(\ref{thaxrh4})] and Table \ref{taeftc}
lead to the results summarised in Table \ref{sumallc}. These values
can be used to calculate the $S$ and $P$ splittings which 
have been measured in the laboratory and their
discrepancy and, eventually, the success of the theory can be evaluated
(Table \ref{tablesplitsn}, Table \ref{tablesplitpc}). We notice
there is a discrepancy of $18\%$ between the theoretical and the
experimental splitting of the
$\chi_{c2}(1P)$ and $\chi_{c1}(1P)$ states. The ratio 
$\rho=(2^{++}-1^{++})/\left(1^{++}-0^{++}\right)$ equals $\rho=0.420+0.005$
which is well below the experimentally calculated ($\equiv 0.49$) 
\cite{a35}. Thus,
although, some of the energy levels seem to be in a very good agreement
with experiment, the results should be taken with great care. The reason
why the charmonium system is not described entirely adequately by the
theoretical model is that our theory considers the static interaction as the
main contributor to the system's energy and it treats the retarded terms
perturbatively. The problem is that for charmonium, this 
scheme does not provide
exactly the correct energy as it nearly does for bottomium, resulting
to larger deviations from experimental values. 
In contrast to bottomium where the corrections due to the
Breit terms constitute
a small proportion of the total energy, in charmonium they are quite
large. Anyway, larger corrections to the static energy (which, also, contains
relativistic
contributions) were expected since these terms are inversely
proportional to the mass of the constituent particles. 
By looking at the expectation value of the square of 
the speed of the
systems (Table \ref{sumall}, Table \ref{sumallaa} and Table \ref{sumallc}), we
can conclude that, indeed, although the relativistic corrections are not
negligible, to some extent, both bottomium and charmonium can be regarded as non-relativistic systems, 
with this assertion suiting more to bottomium. 

Another reason for the failure to obtain the correct results is the
form of the potential that is supposed to describe strong interactions.
Although the potential employed exhibits the correct behaviour at small and
large distances, we are not able to determine what it looks like in
between. In other words, we do not know its exact form. Certainly, a
knowledge of this would lead to better results. 

In addition, we should not forget the order to which the energy 
values has been calculated. If the wavefunction used, corresponded to the
solutions of the Coulomb problem, the spectrum would contain corrections
of order up to $\alpha^4_s$. For the wavefunctions of the harmonic
oscillator, the energies contain terms of order up to $\alpha_s$. If we
wish to calculate the energies more accurately, we need to consider
higher order terms since their inclusion would provide an essential
contribution to the energy levels, however, again, the results should
not agree exactly with experiment. This is because it is not possible to
take into account all diagrams. Thus, we should better compare our
results with other theories. We consider the study conducted by Eichten {\it et al.}
(they, also, used a funnel potential as a possible form of the interaction)
\cite{b20} on charmonium. They fitted the parameters ($\alpha_s=0.39$,
$\kappa=926 \mbox{ }\rm GeV/fm $) and calculated the
energies summarised in Table \ref{otherthetable1}, 
Table \ref{otherthetable2}. The
asterisk (*) refers to the states used to fit the parameters. Their
values are slightly different from those we used because at the time
those calculations were carried out, the experimental data differed from the recent
results. The $P$- states, also, in these tables represent the
centre-of-mass state of the $j=0$, $j=1$, $j=2$ states. 

The coupling constant of the interaction and the QCD scale parameter were calculated in the previous subsection where we studied the bottomium spectrum. One might wonder whether it would be wise to allow the values of these parameters to be derived from the study of charmonium, instead. The problem is that in order to estimate these values, one approximation has been made already by optimising the radial wavefunctions of the three-dimensional harmonic oscillator (in addition to the approximation concerning the form of the potential and the approximation related to the effective interaction). A more rigorous and accurate 
procedure should assume the use of the wavefunctions of the
funnel potential. It is true that charmonium exhibits a more relativistic behaviour than bottomium and a very careful consideration should be taken, otherwise there is a danger of obtaining unreliable results. The fact that bottomium consists of heavier particles and thereby the system is regarded as less relativistic implies that such approximations as those performed throughout this work are less likely to result in unacceptable and meaningless 
results if the parameters are calculated for bottomium.

\section{Conclusions}
\label{sec:ten}

In this paper, the application of the Breit equation to bound state of systems characterised by short-range interactions was studied. Despite the clear lack of covariance of the equation, we assumed that it can be considered to be approximately Lorentz invariant for systems interacting through a very strong short-range potentials such as quarkonia and more specifically, bottomium and charmonium. A funnel potential was introduced as a candidate potential describing QCD and the Pauli approximation of the Breit equation was employed in order to distinguish the contribution of the static terms of the interaction from that part of the interaction responsible for retardation. The solution of the Breit equation helped to obtain the spectrum of bottomium and charmonium and to compare them with experimental data. The $S$ and $P$ states as well as their splitting were calculated and it turned out that the results are, in general, in a very good agreement with experiment. The Breit equation was, also, used to determine the coupling constant and the scale parameter of the strong interaction and they were estimated to be $\alpha_s = 0.36 \pm 0.04$ and $\Lambda = 409 \pm 40 \rm GeV$, respectively.

A part of the success was attributed to the fact that the small speed of the states allowed the Breit terms to be treated perturbatively despite the relatively large coupling constant ($\alpha_s/m^2$ is small while $\alpha_s$ is large). The large mass of the constituent particles makes the systems less relativistic, therefore the static part of the interaction contributes much more to the binding energy than the Breit terms. 

Unfortunately, for systems that are characterised by larger speed or, equivalently, by a smaller mass, the discrepancy between experiment and the theory is large enough to raise concerns over the reliability of the Breit equation as a satisfactory theory. This suggests that, in that case, the Breit equation is not adequate to describe systems of two fermions and an alternative equation should be employed. 

\section{Acknowledgments}
I would like to thank A. Anselm and N. Dombey for their interesting and essential comments on the Breit equation.


\clearpage
\begin{table}[t]
\begin{center}
\begin{tabular}
{|l|l|l|} \hline
Meson & Energy due to  & Energy corrections due to  \\
States & the Static potential &
the Breit terms
$\left(A_{\rm{index}}=\frac{1}{9}\frac{\alpha_s}{m^2_b}\sqrt
{\frac{\lambda_{\rm{index}} ^3}{\pi}}
\right)$ \\ \hline
$\Upsilon (1S)$ & $E_{{\rm static}}(\Upsilon (1S))$ &    $+8.0A_{1S}$ \\
\hline
$\Upsilon (2S)$ & $E_{{\rm static}}(\Upsilon (2S))$ &    $-80.0A_{2S}$
\\
\hline
$\Upsilon (3S)$ & $E_{{\rm static}}(\Upsilon (3S))$ &    $-183.2A_{3S}$
\\
\hline
$\Upsilon (4S)$ & $E_{{\rm static}}(\Upsilon (4S))$ &    $-279.7A_{4S}$
\\
\hline
$\Upsilon (5S)$ & $E_{{\rm static}}(\Upsilon (5S))$ &    $-371.0A_{5S}$
\\
\hline
$\Upsilon (6S)$ & $E_{{\rm static}}(\Upsilon (6S))$ &    $-458.1A_{6S}$
\\
\hline
$\chi_{b0}(1P)$ & $E_{{\rm static}}(\chi_{b0}(1P))$ &    $-320.0A_{1P}$
\\
\hline
$\chi_{b1}(1P)$ & $E_{{\rm static}}(\chi_{b1}(1P))$ &    $-160.0A_{1P}$
\\
\hline
$\chi_{b2}(1P)$ & $E_{{\rm static}}(\chi_{b2}(1P))$ &    $-70.4A_{1P}$
\\
\hline
$\chi_{b0}(2P)$ & $E_{{\rm static}}(\chi_{b0}(2P))$ &    $-518.4A_{2P}$
\\
\hline
$\chi_{b1}(2P)$ & $E_{{\rm static}}(\chi_{b1}(2P))$ &    $-310.4A_{2P}$
\\
\hline
$\chi_{b2}(2P)$ & $E_{{\rm static}}(\chi_{b2}(2P))$ &    $-193.9A_{2P}$
\\
\hline
\end{tabular}
\end{center}
\caption {Energy levels of bottomium as a function of the parameters
$\alpha_s,  m_b, \lambda$s.}

\label{taeft}
\end{table}
\begin{table}[b]
\begin{center}
\begin{tabular}
{|l|l|l|l|} \hline
Meson & $J^{PC}$ &Energy due to  & Energy corrections due to  \\   
States & &the Static potential & 
the Breit terms
$\left(A_{\rm{index}}=\frac{1}{9}\frac{\alpha_s}{m^2_b}\sqrt
{\frac{\lambda_{\rm{index}} ^3}{\pi}}
\right)$ \\ \hline
$\eta(1S)$ & $0^{-+}$  &$E_{{\rm static}}(\eta (1S))$ &    $-48.0A_{1S}$
\\
\hline
$\eta (2S)$ & $0^{-+}$  &$E_{{\rm static}}(\eta(2S))$ &
$-416.0A_{2S}$
\\
\hline
$\eta (3S)$ & $0^{-+}$  &$E_{{\rm static}}(\eta (3S))$ &
$-603.2A_{3S}$
\\
\hline
$\eta (4S)$ & $0^{-+}$  & $E_{{\rm static}}(\eta(4S))$ &
$-769.7A_{4S}$
\\
\hline
$h(1P)$ & $1^{+-}$  & $E_{{\rm static}}(h(1P))$ &    $-128.0A_{1P}$
\\
\hline
$h (2P)$ & $1^{+-}$  & $E_{{\rm static}}(h(2P))$ &    $-268.8A_{2P}$
\\
\hline
\end{tabular}  
\end{center}
\caption {Energy levels of not experimentally
measured bottomimum states as a
function of the parameters
$\alpha_s, m_b, \lambda$s.}
 
\label{taeftmore}
\end{table}   
\clearpage
\begin{table}[t]
\begin{center}
\begin{tabular}
{|l|c|l|c|c|l|} \hline
Meson & $J^{PC}$& 
Energy  &\multicolumn{2}{c|} {\rm Energy \mbox{ }(Theory)(MeV)} 
&  \\   \cline{4-5}
States & &(Experim.)&Static part&Retarded part& 
$\left<\left(\frac{v}{c}\right)^2\right>$\\
  &  & (MeV)  & (MeV) & (MeV) &\\ \hline
$\Upsilon(1S)$  & $1^{--}$& $9460.37\pm 0.21$ & \multicolumn {2}{c|} 
{$9460.37\pm 0.21\mbox{ }^*$} & 0.050 \\ \cline{4-5} 
               2nd & & & $ 9454.89\pm 0.21$ & $+5.48 \pm 0.01$ & \\ \hline 
$\Upsilon(2S)$  & $1^{--}$& $10023.30\pm 0.31$ & \multicolumn {2}{c|}
{$10023.30\pm 0.31\mbox{ }^*$} &  0.083 \\ \cline{4-5}
               8th & & & $ 10027.40\pm 0.31$ & $-4.101 \pm 0.007$ & \\ 
\hline       
$\Upsilon(3S)$  & $1^{--}$& $10355.3\pm 0.5$ & \multicolumn {2}{c|}
{$10355.3\pm 0.5\mbox{ }^*$} &  0.111\\ \cline{4-5}
              14th  & & & $10362.9 \pm 0.5$ & $-7.42 \pm 0.01$ & \\ \hline
$\Upsilon(4S)$  & $1^{--}$& $10580.0\pm 3.5$ & \multicolumn {2}{c|}
{$10580.0\pm 3.5\mbox{ }^*$} &  0.087\\ \cline{4-5}
               16th & & & $ 10584.9\pm 3.5$ & $-4.929 \pm 0.008$ & \\ \hline
$\Upsilon(5S)$  & $1^{--}$& $10865\pm 8$ & \multicolumn {2}{c|}
{$10882\pm 8$} &  0.125\\ \cline{4-5}
               17th & & & $10890 \pm 8$ & $-7.86 \pm 0.01$ & \\ \hline
$\Upsilon(6S)$  & $1^{--}$& $11019\pm 8$ & \multicolumn {2}{c|}
{$11037\pm 9$} &  0.083\\ \cline{4-5}
               18th & & & $11041 \pm 9$ & $ -3.944 \pm 0.007$ & \\ \hline
$\eta(1S)$  & $0^{-+}$&  & \multicolumn {2}{c|}
{$9393.9\pm 0.3$} & 0.050 \\ \cline{4-5}
               1st & & & $ 9426.8\pm 0.3$ & $ -32.88 \pm 0.08 $ & \\ \hline
$\eta(2S)$  & $0^{-+}$&  & \multicolumn {2}{c|}
{$9973.3\pm 0.4$} &  0.083\\ \cline{4-5}
               7th  & & & $9994.6 \pm 0.4$ & $ -21.32\pm 0.04$ & \\ \hline
$\eta(3S)$  & $0^{-+}$&  & \multicolumn {2}{c|}
{$10324\pm 2$} & 0.111 \\ \cline{4-5}
               13th & & & $ 10348\pm 2$ & $-24.42 \pm 0.04$ & \\ \hline
$\eta(4S)$  & $0^{-+}$& & \multicolumn {2}{c|}
{$10553\pm 1$} &  0.087\\ \cline{4-5}
              15th & & & $ 10567\pm 1$ & $ -13.56\pm 0.02$ & \\ \hline
$\chi_{b0}(1P)$  & $0^{++}$& $9859.8\pm 1.3$ & \multicolumn {2}{c|}
{$9858.9\pm 0.4$} & 0.043 \\ \cline{4-5}
               3rd  & & & $ 9869.1\pm 0.4$ & $-10.15 \pm 0.02$ & \\
\hline
$\chi_{b1}(1P)$  & $1^{++}$& $9891.9\pm 0.7$ & \multicolumn {2}{c|}
{$9891.9\pm 0.7\mbox{ }^*$} &  0.043\\ \cline{4-5}
             5th   & & & $9897.0 \pm 0.7$ & $-5.077 \pm 0.001$ & \\
\hline
$\chi_{b2}(1P)$  & $2^{++}$& $9913.2\pm 0.6$ & \multicolumn {2}{c|}
{$9914.7\pm 0.9$} & 0.043 \\ \cline{4-5}
              6th  & & & $ 9916.9 \pm 0.9$ & $-2.234 \pm 0.004$ & \\
\hline
 
$h(1P)$  & $1^{+-}$& $ $ & \multicolumn {2}{c|}
{$9906 \pm 1$} & 0.043 \\ \cline{4-5}
              4th  & & & $ 9910 \pm 1 $ & $-4.062  \pm 0.007$ & \\   
\hline

\end{tabular}
\end{center}
\caption {The energy spectrum of the bottomium system derived from 
experiment (third column), the Breit equation (fourth and fifth 
columns) and the expectation values of the square of 
the velocity operator $v$ for each state (sixth column).
The asterisk (*) refers to the states used to fit the parameters. The
position of every state in the spectrum (for the states we have
considered) is mentioned in the first column.}
 
\label{sumall}
\end{table}
\clearpage
\begin{table}[t]
\begin{center}
\begin{tabular}
{|l|c|l|c|c|l|} \hline
Meson & $J^{PC}$&
Energy  &\multicolumn{2}{c|} {\rm Energy \mbox{ }(Theory)(MeV)}
&  \\   \cline{4-5}
States & &(Experim.)&Static part&Retarded part& $\left<\left
(\frac{v}{c}\right)^2\right>$\\ 
  &  & (MeV)  & (MeV) & (MeV) &\\ \hline
$\chi_{b0}(2P)$  & $0^{++}$& $10232.1\pm 0.6$ & \multicolumn {2}{c|}
{$10234.3\pm 0.2$} & 0.087 \\ \cline{4-5}
               9th & & & $10253.9 \pm 0.2$ & $ -19.60\pm 0.03$ & \\ \hline
$\chi_{b1}(2P)$  & $1^{++}$& $10255.2\pm 0.5$ & \multicolumn {2}{c|}
{$10255.2\pm 0.5\mbox{ }^*$} & 0.087 \\ \cline{4-5}
              11th  & & & $10266.9 \pm 0.5$ & $ -11.74\pm 0.02$ & \\ \hline
$\chi_{b2}(2P)$  & $2^{++}$& $10268.5\pm 0.4$ & \multicolumn {2}{c|}
{$10266.9\pm 0.7$} & 0.087 \\ \cline{4-5}
              12th  & & & $ 10274.3\pm 0.7$ & $ -7.33 \pm 0.01$ & \\ \hline
$h(2P)$  & $1^{+-}$&  & \multicolumn {2}{c|}
{$10260.2\pm 0.9$} & 0.087 \\ \cline{4-5}
               10th & & & $ 10270.4 \pm 0.9$ & $ -10.17 \pm 0.02$ & \\ \hline
\end{tabular}
\end{center}
\caption {The energy spectrum of the bottomium system derived from 
experiment (third column), the Breit equation (fourth and fifth 
columns) and the expectation values of the square of 
the velocity operator $v$ for each state (sixth column).
The asterisk (*) refers to the states used to fit the parameters. The
position of every state in the spectrum (for the states we have
considered) is mentioned in the first column.}
 
\label{sumallaa}
\end{table}

\begin{table}[b]
\begin{center}
\begin{tabular}
{|c|l|l|c|} \hline
$\Delta M$ & Theory (MeV)  & Experiment (MeV) &Deviation (\%) \\
\hline
$m \left(\chi_{b1}(1P)\right)-m \left(\chi_{b0}(1P)\right)$ & $33.0
\pm 0.8$ & $32 \pm 1$ & 3.1\\ \hline
$m \left(\chi_{b2}(1P)\right)-m \left(\chi_{b1}(1P)\right)$ & $22.8
\pm 1.1$ & $21.3 \pm 0.9$ &7.0\\ \hline
$m \left(\chi_{b1}(2P)\right)-m \left(\chi_{b0}(2P)\right)$ & $20.9
\pm 0.8$ & $23.1 \pm 0.8$ & 9.5\\ \hline
$m \left(\chi_{b2}(2P)\right)-m \left(\chi_{b1}(2P)\right)$ & $11.7
\pm 0.9$ & $13.3 \pm 0.6 $ & 12.0
\\ \hline
\end{tabular}
\end{center}
\caption {Splittings of the $1P$ and $2P$ levels of bottomium.}
 
\label{tablesplit}
\end{table}
\clearpage

\begin{table}[t] 
\begin{center}  
\begin{tabular}
{|c|l|} \hline
$\Delta M$ & Theory (MeV)  \\
\hline
$m \left(\Upsilon(1S)\right)-m \left(\eta (1S)\right)$ & $66.5 \pm 0.4$\\ \hline
$m \left(\Upsilon(2S)\right)-m \left(\eta (2S)\right)$ & 
$ 50.0 \pm 0.5 $\\ \hline    
      
$m \left(\Upsilon(3S)\right)-m \left(\eta (3S)\right)$ & $31\pm 2$\\ \hline 
$m \left(\Upsilon(4S)\right)-m \left(\eta (4S)\right)$ &  $27\pm 4$\\ \hline
\end{tabular}
\end{center}
\caption {Splittings of the $S$ levels of bottomium.}
 
\label{tablesplits}
\end{table}

\begin{table}[b]
\begin{center}
\begin{tabular}
{|l|l|l|} \hline
Meson & Energy due to  & Energy corrections due to  \\
States & the Static potential &
the Breit terms
$\left(A_{{\rm index}}=\frac{1}{9}\frac{\alpha_s}{m^2_c}
\sqrt{\frac{\lambda_{{\rm index}}^3}{\pi}}
\right)$ \\ \hline
$J/\psi (1S)$ & $E_{{\rm static}}(J/\psi (1S))$ &    $+8.0A_{1S}$ \\
\hline
$\psi (2S)$ & $E_{{\rm static}}(\psi (2S))$ &    $-80.0A_{2S}$ \\
\hline 
$\psi (3S)$ & $E_{{\rm static}}(\psi (3S))$ &    $-183.2A_{3S}$ \\
\hline
$\psi (4S)$ & $E_{{\rm static}}(\psi (4S))$ &    $-279.7A_{4S}$ \\
\hline
$\eta_c(1S)$ & $E_{{\rm static}}(\eta (1S))$ &    $-48.0A_{1S}$ \\
\hline
$\eta_c(2S)$ & $E_{{\rm static}}(\eta (2S))$ &    $-416.0A_{2S}$ \\
\hline
$\eta_c(3S)$ & $E_{{\rm static}}(\eta (3S))$ &    $-603.2A_{3S}$ \\
\hline 
$\eta_c(4S)$ & $E_{{\rm static}}(\eta (4S))$ &    $-769.7A_{4S}$ \\
\hline 
$h_c(1P)$ & $E_{{\rm static}}(h_c (1S))$ &    $-128.0A_{1S}$ \\
\hline 
$\chi_{c0}(1P)$ & $E_{{\rm static}}(\chi_{c0}(1P))$ &
$-320.0A_{1P}$\\
\hline 
$\chi_{c1}(1P)$ & $E_{{\rm static}}(\chi_{c1}(1P))$ &
$-160.0A_{1P}$\\
\hline
$\chi_{c2}(1P)$ & $E_{{\rm static}}(\chi_{c2}(1P))$ &    $-70.4A_{1P}$\\
\hline
$\chi_{c1}(1P)$ & $E_{{\rm static}}(\chi_{c1}(1P))$ &
$-160.0A_{1P}$\\
\hline
$\chi_{c2}(1P)$ & $E_{{\rm static}}(\chi_{c2}(1P))$ &    $-70.4A_{1P}$\\
\hline

\end{tabular}   
\end{center}
\caption {Energy levels of charmonium as a function of the parameters   
$\alpha_s,  m_c, \mu, \lambda$s.}
\label{taeftc}
\end{table}
\clearpage
\begin{table}[t]
\begin{center}
\begin{tabular}
{|l|c|l|c|c|l|} \hline
Meson & $J^{PC}$&
Energy  &\multicolumn{2}{c|} {\rm Energy \mbox{ }(Theory)(MeV)}
&  \\   \cline{4-5}
States & &(Experim.)&Static part&Retarded part&
$\left<\left(\frac{v}{c}\right)^2\right>$\\
  &  & (MeV)  & (MeV) & (MeV) &\\ \hline
$J/\psi(1S)$  & $1^{--}$& $3096.88\pm 0.04$ & \multicolumn {2}{c|}
{$3096.88\pm 0.04\mbox{ }^*$} & 0.24 \\ \cline{4-5}
               2nd & & & $ 3078.68\pm 0.04$ & $+18.2 \pm 0.1$ & \\
\hline
$\psi(2S)$  & $1^{--}$& $3686.00\pm 0.09$ & \multicolumn {2}{c|}
{$3686.00\pm 0.09\mbox{ }^*$} &  0.47 \\ \cline{4-5}
               8th & & & $ 3703.58\pm 0.08$ & $-17.58 \pm 0.03$ & \\
\hline
$\psi(3S)$  & $1^{--}$& $4040\pm 10$ & \multicolumn {2}{c|}
{$4106\pm 10$} &  0.36\\ \cline{4-5}
              10th  & & & $4120 \pm 10$ & $-13.54 \pm 0.02$ & \\
\hline
$\psi(4S)$  & $1^{--}$& $4415\pm 6$ & \multicolumn {2}{c|}
{$4454\pm 8$} &  0.37\\ \cline{4-5}
               12th & & & $ 4467\pm 8$ & $-13.44 \pm 0.02$ & \\
\hline
$\eta(1S)$  & $0^{-+}$& $2979.8\pm 2.1$ & \multicolumn {2}{c|}
{$2987\pm 2$} & 0.24\\ \cline{4-5}
               1st & & & $ 3095\pm 2$ & $ -108.0 \pm 0.8 $ & \\
\hline
$\eta(2S)$  & $0^{-+}$& $3594 \pm 5 $ & \multicolumn {2}{c|}
{$3601\pm 1$} &  0.47\\ \cline{4-5}
               7th  & & & $3692 \pm 1$ & $ -91.4\pm 0.2$ & \\
\hline
$\eta(3S)$  & $0^{-+}$&  & \multicolumn {2}{c|}
{$4023\pm 1$} & 0.36 \\ \cline{4-5}
               9th & & & $ 4068\pm 1$ & $-44.59 \pm 0.08$ & \\
\hline
$\eta(4S)$  & $0^{-+}$&  & \multicolumn {2}{c|}
{$4379.1\pm 0.9$} & 0.37 \\ \cline{4-5}
               11th & & & $4416.1 \pm 0.9$ & $-36.98\pm 0.07$ & \\ \hline
$\chi_{c0}(1P)$  & $0^{++}$& $3417.3\pm 2.8$ & \multicolumn {2}{c|}
{$3422\pm 1$} & 0.30 \\ \cline{4-5}
               3rd  & & & $ 3481\pm 1$ & $-59.0 \pm 0.1$ & \\
\hline
$\chi_{c1}(1P)$  & $1^{++}$& $3510.53\pm 0.12$ & \multicolumn {2}{c|}
{$3510.53\pm 0.12\mbox{ }^*$} &  0.30\\ \cline{4-5}
             5th   & & & $3540.02 \pm 0.12$ & $-29.49 \pm 0.05$ & \\
\hline
$\chi_{c2}(1P)$  & $2^{++}$& $3556.17\pm 0.13$ & \multicolumn {2}{c|}
{$3547.9\pm 0.2$} & 0.30 \\ \cline{4-5}
              6th  & & & $ 3560.9 \pm 0.2$ & $-12.98 \pm 0.02$ & \\
\hline   
$h(1P)$  & $1^{+-}$& $ 3526.14\pm 0.24$ & \multicolumn {2}{c|}
{$3483 \pm 2$} & 0.30 \\ \cline{4-5}
              4th  & & & $ 3507 \pm 2 $ & $-23.60  \pm 0.04$ & \\ \hline
\end{tabular}
\end{center}
\caption {The energy spectrum of the charmonium system derived from the
experiment (third column), the Breit equation (fourth and fifth 
columns) and the expectation values of the square of
the velocity operator $v$ for each state (sixth column).
The asterisk (*) refers to the states used to fit the parameters. The
position of every state in the spectrum (for the states we have
considered) is mentioned in the first column.}
 
\label{sumallc}
\end{table}  
\clearpage
\begin{table}[t]
\begin{center}
\begin{tabular}
{|c|l|l|c|} \hline
$\Delta M$ & Theory (MeV)  &Experiment (MeV) &Deviation (\%)\\
\hline
$m \left(\psi(1S)\right)-m \left(\eta_c (1S)\right)$ & $110 \pm 2$
& $117\pm 2$ &$6.0$\\ \hline
$m \left(\psi(2S)\right)-m \left(\eta_c (2S)\right)$ &
$ 85 \pm 1 $ & $92\pm 5$& $7.6$\\ \hline
$m \left(\psi(3S)\right)-m \left(\eta_c(3S)\right)$ & $83\pm 10$& &\\
\hline
$m \left(\psi(4S)\right)-m \left(\eta_c (4S)\right)$ &  $75\pm 8$& &\\
\hline
\end{tabular}
\end{center}
\caption {Splittings of the $S$ levels of charmonium.}
 
\label{tablesplitsn}
\end{table}

\begin{table}[b] 
\begin{center}
\begin{tabular}
{|c|l|l|c|} \hline
$\Delta M$ & Theory (MeV)  & Experiment (MeV) &Deviation (\%) \\
\hline
$m \left(\chi_{c1}(1P)\right)-m \left(\chi_{c0}(1P)\right)$ & $89
\pm 1$ & $96 \pm 1$ & 4.3\\ \hline
$m \left(\chi_{c2}(1P)\right)-m \left(\chi_{c1}(1P)\right)$ & $37.4
\pm 0.2$ & $45.6 \pm 0.2$ &18.0\\ \hline
\end{tabular}
\end{center}
\caption {Splittings of the $1P$ states of charmonium.}
 
\label{tablesplitpc}
\end{table}
\clearpage
\begin{table}[t]
\begin{center}
\begin{tabular}
{|c|c|} \hline
\multicolumn{2}{|c|}{Charmonium $m_c=1.84\mbox{ }\rm GeV$} \\ \hline
State & Energy (MeV)\\ \hline
$J/\psi(1S)$      &    $  3095^*$ \\ \hline 
$\psi(2S)$      &    $  3684^*$ \\ \hline
$\psi(3S)$      &    $  4110$ \\ \hline
$\psi(4S)$      &    $  4460$ \\ \hline 
$\psi(5S)$      &    $  4790$ \\ \hline
$\chi_c(1P)$     &    $3522^*$      \\ \hline
\end{tabular}
\end{center}   
\caption{The charmonium spectrum according to Eichten {\sl et al.}
\cite{b20}}
\label{otherthetable1}
\end{table}

\begin{table}[b]
\begin{center}
\begin{tabular}
{|c|c|} \hline
\multicolumn{2}{|c|}{Bottomium $m_b=5.17\mbox{ }\rm GeV$} \\ \hline
State & Energy (MeV)\\ \hline
$\Upsilon(1S)$      &    $  9460^*$ \\ \hline 
$\Upsilon(2S)$      &    $  10050$ \\ \hline
$\Upsilon(3S)$      &    $  10400$ \\ \hline
$\Upsilon(4S)$      &    $  10670$ \\ \hline 
$\Upsilon(5S)$      &    $  10920$ \\ \hline
$\Upsilon(6S)$      &    $  11140$ \\ \hline
$\chi_b(1P)$        &    $   9960$      \\ \hline
$\chi_b(2P)$        &    $  10310$      \\ \hline
$\chi_b(3P)$        &    $  10600$      \\ \hline
\end{tabular}
\end{center}   
\caption{The bottomium spectrum according to Eichten {\sl et al.} \cite{b20}}
\label{otherthetable2}
\end{table}

\end{document}